\documentclass[12pt]{article}
\usepackage{cite,graphicx,amsmath,amssymb}
\usepackage{graphicx}
\usepackage{comment}
\usepackage{multicol}
\usepackage{xcolor}
\usepackage{slashed}
\usepackage[normalem]{ulem}
\usepackage{bbm}
\usepackage{hyperref}
\usepackage{physics}
\usepackage{amsmath}
\usepackage{tikz}
\usepackage{mathdots}
\usepackage{yhmath}
\usepackage{cancel}
\usepackage{color}
\usepackage{siunitx}
\usepackage{array}
\usepackage{multirow}
\usepackage{amssymb}
\usepackage{tabularx}
\usepackage{extarrows}
\usepackage{booktabs}
\usetikzlibrary{fadings}
\usetikzlibrary{patterns}
\usetikzlibrary{shadows.blur}
\usetikzlibrary{shapes}

\usepackage{makecell}

\usepackage{float}
\usepackage{cancel}
\usepackage{longtable}

\newcommand{\be}{\begin{equation}}
\newcommand{\ee}{\end{equation}}
\newcommand{\bea}{\begin{eqnarray}}
\newcommand{\eea}{\end{eqnarray}}
\newcommand*{\hdash}[1][2em]{\rule[0.5ex]{#1}{0.55pt}}
\newcommand*{\fdash}[1][5.5em]{\rule[0.5ex]{#1}{0.55pt}}
\usepackage{mdframed}
\mdfdefinestyle{statementstyle}{
   frametitlerule = true,
   frametitlefont = {\normalfont\bfseries},
   frametitlebackgroundcolor = blue!10,
}
\mdtheorem[style=statementstyle]{statement}{Compactness of brane moduli space}
\begin{document}

\begin{center}

{\Large\bf Compactness of Brane Moduli and the \\ String Lamppost Principle in $d>6$}
\\[2cm]

{\large Alek Bedroya, Yuta Hamada, Miguel Montero and Cumrun Vafa}

\vspace{1cm}

{\it
Jefferson Physical Laboratory, Harvard University, Cambridge, MA 02138, USA
}

\vspace{.3cm}
\large{\today}\\[1.6cm]
%\large{May 24, 2021}

{\bf Abstract}
\flushleft
We demonstrate the validity of the String Lamppost Principle -- that all consistent theories of quantum gravity are in the String Landscape -- for supersymmetric theories in $d>6$ using compactness and connectedness of the moduli space of small instantons, as well as the classification of the associated Coulomb branch.
\end{center}

\newpage
\tableofcontents 
\newpage

\section{Introduction}

Many of the special features of the String Landscape seem to emerge from the exceptional structure of the compactification geometry, which a priori seems inaccessible to a low energy observer.   These are, in turn, directly related to many of the conjectures made in the context of the Swampland program \cite{Vafa:2005ui}.  Thus one path to establishing the Swampland criteria would be to somehow make the internal string geometry ``visible'' to the low energy observer.  This may sound like an impossible task, as it may appear we need access to the KK modes to ``see'' the internal geometry.

 However, this idea was recently accomplished for 8d supersymmetric theories in \cite{Hamada:2021bbz} by studying brane probes and using black hole physics to argue for the compactness of their moduli space which in turn leads to the internal F-theory geometry \cite{Vafa:1996xn,Morrison:1996na,Morrison:1996pp}. 
The main aim of this work is to extend this to the study of all supersymmetric theories in $d\geq 6$.  In particular, the combination of compactness of the moduli of supersymmetric brane probe as well as the structure of CFT's which arise for the probes associated with gauging instantons is a powerful tool in classifying all the possible consistent quantum gravity theories with supersymmetry.  We complete this program, and show that the String Lamppost Principle (SLP) holds for supersymmetric theories in $d>7$.   For theories with maximal supersymmetry, there is not much to show as the low energy theory is fixed by supersymmetry.  For theories with minimal supersymmetry, in the cases in $d\leq 9$ this was not known before.  The brane probe that we consider is the gauge instanton probe or equivalently the $(d-5)$-brane probe, which is the magnetic dual object of the 1-brane probe.  We argue that their moduli are connected through their Coulomb branch.  Moreover, the Coulomb branch should include CFT points arising from the gauge instantons.  In this way, we get a handle on what gauge groups can occur in the theory.  In $d=9$ we use particular facts of 5d SCFT's, as was done in the case of $d=8$ for 4d SCFT's in \cite{Hamada:2021bbz}.  In particular we explain why in the $d=9$ case the $Sp(n)$ gauge symmetry is not part of any quantum gravity landscape (for $n>1$),\footnote{The notation $Sp(1)=SU(2)$ is used.} despite the fact that there is nothing a priori wrong with the corresponding supersymmetric field theory without gravity and is indeed realized in string theory (up to a ``stringy Landau pole'').  
We extend the work of  \cite{Hamada:2021bbz} in $d=8$ to also complete the SLP in that dimension.
 For $d=7$, we show this also completes the SLP modulo a few facts which has yet to be established for 3d ${\cal N}=4$ SCFT's.
We also discuss the SLP in $d=6$. 

The organization of this paper is as follows: In Section \ref{GD} we review the general setup for this paper.  In Section \ref{Sec:constraints} we apply the general setup to various dimensions $6\leq d\leq 9$.  Finally in Section \ref{Sec:conclusion} we close with some concluding thoughts.

\section{General discussion}\label{GD}

We begin with a general outline and discussion of the basic idea underlying this project, which was initiated in \cite{Hamada:2021bbz}. A consistent quantum theory of gravity often includes long-range gauge forces as part of its low-energy description. This is particularly true of supersymmetric theories, where these are often required by supersymmetry. For instance, in theories with sixteen supercharges, the gravity multiplet includes a massless 2-form field. Completeness of the spectrum \cite{Polchinski:2003bq} then requires that all possible values of the charges are populated by physical objects in the theory. Studying the worldvolume theory of these branes, which map in the context of the String Landscape to ``probe branes'' has been the source of much recent progress \cite{Kim:2019vuc,Lee:2019skh,Kim:2019ths,Katz:2020ewz,Hamada:2021bbz,Tarazi:2021duw} and can be used to banish to the Swampland some naively consistent theories of quantum gravity that do not appear in the String Landscape. 

When the branes are supersymmetric, they often have exact moduli spaces -- exactly massless directions of their worldvolume field theory. Some of these moduli have a direct spacetime interpretation, like the scalars parametrizing the ``center of mass'' degrees of freedom of the object in spacetime. But others characterize purely internal degrees of freedom. For instance, in theories of quantum gravity arising via compactification from a higher-dimensional theory, we will often have additional scalars parametrizing the position of the branes in the compact space. Of course, scalars can also have other interpretations, such as Wilson lines of higher-dimensional gauge fields or more exotic origins. In any case, the low-energy effective field theory controlling the dynamics of the moduli is simply a sigma model from the brane worldvolume to the moduli space $\mathcal{M}$:
\begin{equation}\mathcal{L}\supset\int_{\text{brane}} \sqrt{-g} \frac{G_{IJ}}{2} \partial_\mu\phi^I \partial_\nu \phi^J,\label{probeBlag}\end{equation}
where greek indices run over brane's worldvolume directions, $\phi$ are the moduli, and Latin uppercase indices live in the tangent bundle of $\mathcal{M}$. To understand the dynamics of \eqref{probeBlag}, it is often useful to compactify all spatial worldvolume directions of the $p$-brane on the torus $T^p$ to obtain an effective quantum mechanics with target space $\mathcal{M}$ (times any additional degrees of freedom that may arise due to compactification, such as Wilson lines, etc.).   Here we assume $p<d-2$, so the resulting 0-brane has codimension more than 2 in the uncompactified spacetime.  Canonical quantization then produces a spectrum whose energies are equal to the eigenvalues of the Laplacian on $\mathcal{M}$.  We reach the conclusion that the sigma model \eqref{probeBlag} has a spectrum dictated by the Laplacian on $\mathcal{M}$ (times any additional space). In particular, if $\mathcal{M}$ is non-compact and the Laplacian has a continuous spectrum, the set of asymptotic states of the theory described in \eqref{probeBlag} has infinitely many modes of any given finite energy range.  In particular, this means that the entropy density is infinite. Such behavior is fine in quantum field theory, but it is unacceptable in a quantum theory of gravity, where the entropy is upper bounded by that of the corresponding Schwarzschild's black hole or black brane, as specified by Bekenstein's bound. Thus, the consistency of quantum gravity leads us to the basic claim \cite{Hamada:2021bbz} which is the main tool we use in this paper:

\begin{statement*}The moduli space of any $p$-brane with $p<d-2$ is compact (or more precisely has a discrete spectrum of the Laplacian) in a consistent quantum theory of gravity.\end{statement*}

Throughout this paper, we will apply this principle to the magnetic $(d-5)$-brane associated with the magnetic dual of the $B$ field in the gravity multiplet of $d$-dimensional theory with 16 supercharges, extending the results in \cite{Hamada:2021bbz}. These branes preserve half the supercharges in their worldvolume, and when the $d$-dimensional moduli are tuned such that the $d$-dimensional theory has non-Abelian gauge fields, the $(d-5)$-branes correspond to the zero-size limit of gauge theory instantons. An example is the heterotic NS5 brane in 10 dimensions \cite{Witten:1995gx,Ganor:1996mu,Seiberg:1996vs}.

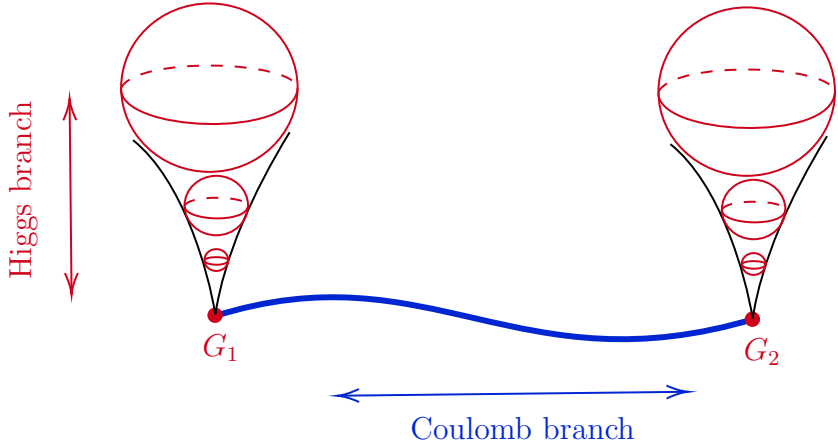
\begin{figure}
    \centering

\tikzset{every picture/.style={line width=0.75pt}} %set default line width to 0.75pt        

\begin{tikzpicture}[x=0.75pt,y=0.75pt,yscale=-1,xscale=1]
%uncomment if require: \path (0,259); %set diagram left start at 0, and has height of 259

%Curve Lines [id:da11545185445230843] 
\draw [color={rgb, 255:red, 0; green, 38; blue, 209 }  ,draw opacity=1 ][line width=2.25]    (225.88,168.75) .. controls (334.38,134.75) and (378.5,206) .. (498.5,170) ;
%Shape: Circle [id:dp32747459580307314] 
\draw  [color={rgb, 255:red, 208; green, 2; blue, 27 }  ,draw opacity=1 ][fill={rgb, 255:red, 208; green, 2; blue, 27 }  ,fill opacity=1 ] (222.5,168.38) .. controls (222.5,170.24) and (224.01,171.75) .. (225.88,171.75) .. controls (227.74,171.75) and (229.25,170.24) .. (229.25,168.38) .. controls (229.25,166.51) and (227.74,165) .. (225.88,165) .. controls (224.01,165) and (222.5,166.51) .. (222.5,168.38) -- cycle ;
%Curve Lines [id:da7107862119292274] 
\draw    (184.5,80) .. controls (209.5,100) and (222.5,146) .. (226,168) ;
%Curve Lines [id:da0567343821404267] 
\draw    (263.5,76) .. controls (244.5,107) and (230.5,139) .. (226,168) ;
%Shape: Ellipse [id:dp03767628122566524] 
\draw  [color={rgb, 255:red, 208; green, 2; blue, 27 }  ,draw opacity=1 ] (178.5,53.5) .. controls (178.5,30.03) and (198.43,11) .. (223,11) .. controls (247.58,11) and (267.5,30.03) .. (267.5,53.5) .. controls (267.5,76.97) and (247.58,96) .. (223,96) .. controls (198.43,96) and (178.5,76.97) .. (178.5,53.5) -- cycle ;
%Curve Lines [id:da36284834823989076] 
\draw [color={rgb, 255:red, 208; green, 2; blue, 27 }  ,draw opacity=1 ]   (178.5,53.5) .. controls (176.31,75.81) and (266.4,80.06) .. (267.5,53.5) ;
%Curve Lines [id:da021251715818693118] 
\draw [color={rgb, 255:red, 208; green, 2; blue, 27 }  ,draw opacity=1 ] [dash pattern={on 4.5pt off 4.5pt}]  (178.5,53.5) .. controls (187.29,37.56) and (264.2,39.69) .. (267.5,53.5) ;

%Shape: Ellipse [id:dp48180552349353856] 
\draw  [color={rgb, 255:red, 208; green, 2; blue, 27 }  ,draw opacity=1 ] (210.48,113) .. controls (210.48,104.72) and (217.65,98) .. (226.49,98) .. controls (235.33,98) and (242.5,104.72) .. (242.5,113) .. controls (242.5,121.28) and (235.33,128) .. (226.49,128) .. controls (217.65,128) and (210.48,121.28) .. (210.48,113) -- cycle ;
%Curve Lines [id:da7430294156842774] 
\draw [color={rgb, 255:red, 208; green, 2; blue, 27 }  ,draw opacity=1 ]   (210.48,113) .. controls (209.69,120.88) and (242.1,122.38) .. (242.5,113) ;
%Curve Lines [id:da0725567154364013] 
\draw [color={rgb, 255:red, 208; green, 2; blue, 27 }  ,draw opacity=1 ] [dash pattern={on 4.5pt off 4.5pt}]  (210.48,113) .. controls (213.64,107.38) and (241.31,108.12) .. (242.5,113) ;

%Shape: Ellipse [id:dp801618550220107] 
\draw  [color={rgb, 255:red, 208; green, 2; blue, 27 }  ,draw opacity=1 ] (220.51,140.5) .. controls (220.51,137.46) and (223.19,135) .. (226.5,135) .. controls (229.81,135) and (232.5,137.46) .. (232.5,140.5) .. controls (232.5,143.54) and (229.81,146) .. (226.5,146) .. controls (223.19,146) and (220.51,143.54) .. (220.51,140.5) -- cycle ;
%Curve Lines [id:da9638213552035411] 
\draw [color={rgb, 255:red, 208; green, 2; blue, 27 }  ,draw opacity=1 ]   (220.51,140.5) .. controls (220.21,143.39) and (232.35,143.94) .. (232.5,140.5) ;

%Curve Lines [id:da5737667287699106] 
\draw [color={rgb, 255:red, 208; green, 2; blue, 27 }  ,draw opacity=1 ]   (220.51,140.5) .. controls (221.69,138.44) and (232.06,138.71) .. (232.5,140.5) ;

%Shape: Circle [id:dp09574771196458265] 
\draw  [color={rgb, 255:red, 208; green, 2; blue, 27 }  ,draw opacity=1 ][fill={rgb, 255:red, 208; green, 2; blue, 27 }  ,fill opacity=1 ] (493.5,170.38) .. controls (493.5,172.24) and (495.01,173.75) .. (496.88,173.75) .. controls (498.74,173.75) and (500.25,172.24) .. (500.25,170.38) .. controls (500.25,168.51) and (498.74,167) .. (496.88,167) .. controls (495.01,167) and (493.5,168.51) .. (493.5,170.38) -- cycle ;
%Curve Lines [id:da5154009739218925] 
\draw    (455.5,82) .. controls (480.5,102) and (493.5,148) .. (497,170) ;
%Curve Lines [id:da6699141302756757] 
\draw    (534.5,78) .. controls (515.5,109) and (501.5,141) .. (497,170) ;
%Shape: Ellipse [id:dp6819835655760227] 
\draw  [color={rgb, 255:red, 208; green, 2; blue, 27 }  ,draw opacity=1 ] (449.5,55.5) .. controls (449.5,32.03) and (469.43,13) .. (494,13) .. controls (518.58,13) and (538.5,32.03) .. (538.5,55.5) .. controls (538.5,78.97) and (518.58,98) .. (494,98) .. controls (469.43,98) and (449.5,78.97) .. (449.5,55.5) -- cycle ;
%Curve Lines [id:da2852837039105951] 
\draw [color={rgb, 255:red, 208; green, 2; blue, 27 }  ,draw opacity=1 ]   (449.5,55.5) .. controls (447.31,77.81) and (537.4,82.06) .. (538.5,55.5) ;
%Curve Lines [id:da16336796353456506] 
\draw [color={rgb, 255:red, 208; green, 2; blue, 27 }  ,draw opacity=1 ] [dash pattern={on 4.5pt off 4.5pt}]  (449.5,55.5) .. controls (458.29,39.56) and (535.2,41.69) .. (538.5,55.5) ;

%Shape: Ellipse [id:dp3567442385516222] 
\draw  [color={rgb, 255:red, 208; green, 2; blue, 27 }  ,draw opacity=1 ] (481.48,115) .. controls (481.48,106.72) and (488.65,100) .. (497.49,100) .. controls (506.33,100) and (513.5,106.72) .. (513.5,115) .. controls (513.5,123.28) and (506.33,130) .. (497.49,130) .. controls (488.65,130) and (481.48,123.28) .. (481.48,115) -- cycle ;
%Curve Lines [id:da3780753444098368] 
\draw [color={rgb, 255:red, 208; green, 2; blue, 27 }  ,draw opacity=1 ]   (481.48,115) .. controls (480.69,122.88) and (513.1,124.38) .. (513.5,115) ;
%Curve Lines [id:da37999967991647976] 
\draw [color={rgb, 255:red, 208; green, 2; blue, 27 }  ,draw opacity=1 ] [dash pattern={on 4.5pt off 4.5pt}]  (481.48,115) .. controls (484.64,109.38) and (512.31,110.12) .. (513.5,115) ;

%Shape: Ellipse [id:dp6094484291927702] 
\draw  [color={rgb, 255:red, 208; green, 2; blue, 27 }  ,draw opacity=1 ] (491.51,142.5) .. controls (491.51,139.46) and (494.19,137) .. (497.5,137) .. controls (500.81,137) and (503.5,139.46) .. (503.5,142.5) .. controls (503.5,145.54) and (500.81,148) .. (497.5,148) .. controls (494.19,148) and (491.51,145.54) .. (491.51,142.5) -- cycle ;
%Curve Lines [id:da6522642838716841] 
\draw [color={rgb, 255:red, 208; green, 2; blue, 27 }  ,draw opacity=1 ]   (491.51,142.5) .. controls (491.21,145.39) and (503.35,145.94) .. (503.5,142.5) ;

%Curve Lines [id:da5315789817461158] 
\draw [color={rgb, 255:red, 208; green, 2; blue, 27 }  ,draw opacity=1 ]   (491.51,142.5) .. controls (492.69,140.44) and (503.06,140.71) .. (503.5,142.5) ;

%Straight Lines [id:da9732269871069741] 
\draw [color={rgb, 255:red, 0; green, 38; blue, 209 }  ,draw opacity=1 ]   (289.5,208.98) -- (462.5,207.02) ;
\draw [shift={(464.5,207)}, rotate = 539.35] [color={rgb, 255:red, 0; green, 38; blue, 209 }  ,draw opacity=1 ][line width=0.75]    (10.93,-3.29) .. controls (6.95,-1.4) and (3.31,-0.3) .. (0,0) .. controls (3.31,0.3) and (6.95,1.4) .. (10.93,3.29)   ;
\draw [shift={(287.5,209)}, rotate = 359.35] [color={rgb, 255:red, 0; green, 38; blue, 209 }  ,draw opacity=1 ][line width=0.75]    (10.93,-3.29) .. controls (6.95,-1.4) and (3.31,-0.3) .. (0,0) .. controls (3.31,0.3) and (6.95,1.4) .. (10.93,3.29)   ;

%Straight Lines [id:da37563081276554633] 
\draw [color={rgb, 255:red, 208; green, 2; blue, 27 }  ,draw opacity=1 ]   (153.48,156) -- (152.52,61) ;
\draw [shift={(152.5,59)}, rotate = 449.42] [color={rgb, 255:red, 208; green, 2; blue, 27 }  ,draw opacity=1 ][line width=0.75]    (10.93,-3.29) .. controls (6.95,-1.4) and (3.31,-0.3) .. (0,0) .. controls (3.31,0.3) and (6.95,1.4) .. (10.93,3.29)   ;
\draw [shift={(153.5,158)}, rotate = 269.42] [color={rgb, 255:red, 208; green, 2; blue, 27 }  ,draw opacity=1 ][line width=0.75]    (10.93,-3.29) .. controls (6.95,-1.4) and (3.31,-0.3) .. (0,0) .. controls (3.31,0.3) and (6.95,1.4) .. (10.93,3.29)   ;

% Text Node
\draw (120.5,151.5) node [anchor=north west][inner sep=0.75pt]  [rotate=-270] [align=left] {\textcolor[rgb]{0.82,0.01,0.11}{Higgs branch}};
% Text Node
\draw (323,218) node [anchor=north west][inner sep=0.75pt]  [color={rgb, 255:red, 0; green, 38; blue, 209 }  ,opacity=1 ] [align=left] {Coulomb branch};
% Text Node
\draw (218,176.4) node [anchor=north west][inner sep=0.75pt]    {$\textcolor[rgb]{0.82,0.01,0.11}{G_{1}}$};
% Text Node
\draw (491,178.4) node [anchor=north west][inner sep=0.75pt]    {$\textcolor[rgb]{0.82,0.01,0.11}{G}\textcolor[rgb]{0.82,0.01,0.11}{_{2}}$};

\end{tikzpicture}

    \caption{The above figure shows a path between two instantons in the instanton moduli space. The path connects a $G_1$ instanton to a $G_2$ instanton where $G_1$ and $G_2$ are independent non-Abelian components of the spacetime gauge group $G$. We first shrink the $G_1$ instanton down to zero-size, which corresponds to moving along its Higgs branch to the Coulomb branch. Then we move in the Coulomb branch to deform a $G_1$ zero-size instanton to a $G_2$ zero-size instanton. Finally, we can move in the $G_2$ instanton Higgs branch by increasing the size of the instanton.}
    \label{instanton}
\end{figure}

When the instanton is of finite size, larger than the cutoff of the low-energy supergravity description, the low-energy dynamics can be read off from the set of zero modes of supergravity fields in the instanton background. This can be efficiently analyzed using supersymmetry and the index theorem to obtain the number of fermion zero modes. The corresponding theory, including the modulus $\rho$ that parametrizes the instanton size, is known as the ``Higgs branch'' of the brane worldvolume theory. It connects to another branch of the moduli space, the ``Coulomb branch'', at $\rho=0$ (Figure \ref{instanton}). The Higgs branch receives its name because the low-energy effective field theory description there does not contain worldvolume gauge fields, while at low energies on the Coulomb branch, the low-energy effective description is a supersymmetric gauge theory. The phase transition between the two at $\rho=0$ is described by a (possibly free) SCFT. While the Higgs branch can be described via supergravity, the small instanton SCFT and the Coulomb branch cannot.

 The basic question about the Coulomb branch of the theory is its dimension, known as the rank of the theory (see \cite{Cecotti:2021ouq} for a recent review). Even though this cannot be addressed from bulk supergravity, the dimension of the Coulomb branch for a small instanton can be accessed by supersymmetry due to constructions like ADHM for classical groups or Minahan-Nemeschansky theories \cite{Minahan:1996fg,Minahan:1996cj} for exceptional groups. Specifically, the ADHM construction for classical groups allows one to parametrize the low-energy dynamics in terms of linear degrees of freedom (which gives a natural description of the low-energy dynamics of zero-size instantons, see \cite{Witten:1994tz}) and includes one scalar parametrizing the Coulomb branch. As a result, for branes that can arise as a small instanton limit, the Coulomb branch will be rank one\footnote{A priori one may think that this does not exclude the possibility that, e.g., an exotic small $E_8$ instanton can be described by a yet to be discovered SCFT which has a rank higher than 1.  However, since there is a unique moduli space of $E_8$ instantons, studying the small $E_8$ moduli space, using string theory is perfectly allowed to deduce that its Coulomb branch is one dimensional.  This is a local statement that does not rely on the existence of quantum gravity in particular.   We will provide another supporting argument in Section \ref{Sec:8d}.}. 
 
 Although it never happens in known string constructions, a priori, we may also consider the case where a given brane never arises as a small instanton of a non-Abelian group. In this case, the non-commutative geometry version of the ADHM construction works for $U(1)$ instantons in non-commutative space \cite{Hamanaka:2013vca}, and again yields a one-dimensional Coulomb branch.
 
We emphasize that the argument outlined there essentially is that there is a unique field theory description of the moduli space of instantons, including zero-size configurations, that predicts a one-dimensional Coulomb branch. This and its exceptional versions are {\it pure field theory phenomena}, even though some were first discovered in the context of string backgrounds, that we use as building blocks in the construction of quantum theories of gravity below.

The moduli space theory is not only rank one: it is also connected. In \cite{Hamada:2021bbz}, this was argued via a strengthened version of the Cobordism Conjecture \cite{McNamara:2019rup} which was argued there to hold for theories with 8 supercharges because there is no superpotential is allowed for scalar fields\footnote{The only way scalar fields pick up mass in theories with eight supercharge is via coupling to vector multiplets, as in going to their Coulomb branch, and not through self-interaction of the scalar multiplets. See also the discussion in \cite{Cecotti:2021cvv}.}.

 Connectedness of the moduli space refers to instantons of a fixed charge -- components of different charge are obviously disconnected --. In some string compactifications, like the rank 10 theory in eight dimensions, where one can have symplectic groups where the small instanton has a zero-dimensional Coulomb branch. This corresponds to fractional $D3$ branes stuck at $O7^+$ planes. However, crucially, there is never more than one non-Abelian factor with a zero-dimensional Coulomb branch\footnote{A redundant Swampland prediction coming out of this picture is that there can never be a point in moduli space with more than one $Sp(n)$ factor in the rank 10 8d theory. As was shown in \cite{Hamada:2021bbz}, this prediction is correct.}. In the stringy description, this is due to the fact that we have a single $O7^+$ plane. The corresponding moduli space is therefore a single point, which is connected. For these theories, we consider the more interesting instanton moduli space with an instanton number of two, where the Coulomb branch is again one-dimensional.

Finally, we also note that connectedness of the moduli space for the small instanton limit of every possible non-Abelian gauge group with eight supercharges is deeply connected with the fact that they all carry the \emph{same} physical brane charge. As discussed in \cite{Heidenreich:2020pkc}, this is related to demanding the absence of Chern-Weil global symmetries.

To sum up, we end up with the conclusion that the space of the $(d-5)$-branes in theories with 16 supercharges is a connected moduli space, corresponding to a rank one Coulomb branch. The basic consistency principle that will allow us to fully classify these Coulomb branches is just the simple fact that brane worldvolume couplings should be well defined on moduli space, up to duality transformations.

This seemingly mild principle will turn out to have far-reaching consequences, to the extent that we can determine the full moduli space of theories with sixteen supercharges in seven and higher dimensions! 
 
\section{Swampland constraints in various dimensions}\label{Sec:constraints}
In this Section, we consider rank one worldvolume theories with 8 supercharges that describe codimension-4 small instantons in various spacetime dimensions.
By imposing consistency conditions on worldvolume theories, we derive new swampland constraints and are even able to reconstruct, purely from the brane perspective, the internal geometries that are familiar from string theory.

\subsection{9d}\label{Sec:9d}

In this Section, we show that the SLP holds for 9-dimensional supergravity theories.

The gauge instantons are 4-branes, which are described by 5d $\mathcal{N}=1$ theory.
The consistency conditions on the brane theory impose strong restrictions on the gauge algebras.
In particular, we argue that theories with $\mathfrak{sp}(n)$ gauge symmetry are in the swampland for $n>1$ with dynamical quantum gravity, but fine without gravity.
We also reconstruct the internal space $S^1/\mathbb{Z}_2$ of type $I'$ string theory~\cite{Polchinski:1995df} from the viewpoint of the 4-brane.

\subsubsection*{Consistency condition from 4-brane}
We consider a 5d $\mathcal{N}=1$ rank one theory as a worldvolume theory of 4-brane.
This theory has a Coulomb branch of real dimension one.
The Coulomb branch is parametrized by the expectation value of the real scalar field $\phi$ belonging to the vector multiplet $\mathcal{A}$.
The gauge symmetry is $U(1)$ on a generic point of the Coulomb branch.

The 9d theory with sixteen supercharges has gauge symmetry, with propagating gauge bosons. This spacetime gauge symmetry is seen as a global symmetry of the worldvolume theory of the brane.
If there are non-Abelian factors in the spacetime gauge group, there is a point of symmetry enhancement at the Coulomb branch in the worldvolume theory. At this special point, the Coulomb branch connects to a Higgs branch, in which the brane ``fattens up'' as an instanton of the spacetime gauge group.

As discussed in the previous Section, the Coulomb branch moduli space must be compact and connected.
The compactness of the moduli space is required by the finiteness of the black hole entropy, and the connectedness is required by the stronger cobordism conjecture.
Therefore, the Coulomb branch moduli space is connected, one dimensional, and compact, and so it is either $S^1$ or $S^1/\mathbb{Z}_2$.

We will now argue that the case of 16 supercharges we are interested in corresponds to $S^1/\mathbb{Z}_2$, while the $S^1$ moduli space corresponds to nine-dimensional theories with 32 supercharges. Take the 9d theory and compactify on $T^2$, to obtain an $\mathcal{N}=1$ eight-dimensional theory. The instanton moduli space is now complex one-dimensional, with the additional scalar coming from the Wilson line of the photon. In compactifications coming from theories where the instanton Coulomb branch in 9d is $S^1/\mathbb{Z}_2$, the 8d geometry is an elliptic K3, and in particular, it has curvature. By contrast, if the 9d instanton Coulomb branch is $S^1$, the 8d instanton Coulomb branch has geometry $T^2$, which is flat. 

String theory makes a prediction for the possible Coulomb branches of the probe brane worldvolume theory, and these can be studied by analyzing noncompact configurations of branes in string theory. The stringy prediction for the moduli space geometry from this noncompact analysis is always that the moduli space geometry is not flat and has singularities at a finite distance (consider, for instance, a $D3$ probing a $D7$). We will take this prediction that string theory makes for field theory as an assumption; from this, it follows that the moduli space geometry can only be K3, which uplifts to $S^1/\mathbb{Z}_2$ in nine dimensions.

%---------------------
We can provide another heuristic argument for the same conclusion, which may have wider applicability than the current context. From the brane point of view, the basic difference between $S^1$ and an interval is that the former has an isometry\footnote{As one can see by the arguments later in the Section, there cannot be any points on the moduli space where the 4-brane theory corresponds to a gauge instanton when the moduli space is $S^1$.  So indeed, the Coulomb branch has no special points, and it will automatically have continuous shift symmetry, at least in the IR.}. The Coulomb branch parameter $\phi$ is then actually an axion with a continuous shift symmetry. This means that the currents
\begin{equation} J_1=d\phi,\quad J_4=*d\phi\end{equation}
are exactly conserved, $d*J_1=d*J_4=0$, and generate a 0-form and a 4-form global symmetry on the brane worldvolume. The object charged under the 4-form current is simply solitonic membranes of $\phi$, i.e. field configurations $\phi(x_5)$ that depend nontrivially on one spatial coordinate transverse to the membrane and that wind around the target space circle once.

Because we have these symmetries, we can include topological couplings in the worldvolume brane action,
\begin{equation} \int *J_1 \wedge A_1,\quad \int *J_4 \wedge A_4,\end{equation}
which introduce a coupling to the corresponding background connections. This is a standard procedure in field theory. We will now give an argument that, whenever this happens in the worldvolume theory of a brane coupled to quantum gravity, the background connections \emph{must} correspond to dynamical fields of the bulk theory. Otherwise, the worldvolume symmetry becomes an exact global symmetry, which is forbidden in quantum gravity. We can argue for this in a manner similar to Sen's construction of branes within branes via tachyon condensation \cite{Sen:1999nx}. Consider a brane- antibrane pair, with a worldvolume charged state on the brane. The two branes condense, but the condensation cannot be complete since otherwise, the global charge on the worldvolume theory would be violated. More concretely, the winding of the worldvolume scalar forces the tachyon condensation to remain incomplete in an appropriate locus. As a result, one is left with a remnant solitonic object in the bulk spacetime. Whichever process can make this object break or decay would uplift to the original field theory, contradicting the assumption that the theory had a global symmetry.

In the case under consideration, we would therefore conclude that the bulk theory has a 4-form field. Such a field (or rather, its dual 3-form) is part of the 9d $\mathcal{N}=2$ supergravity multiplet, but not part of the 9d $\mathcal{N}=1$ multiplet. As a consequence, $S^1$ is only compatible with 32 supercharges, as advertised. 

An important caveat is that this argument only applies to \emph{exact} symmetries of the worldvolume brane theory. One could have worldvolume accidental IR symmetries, which will not be coupled to a dynamical bulk field. An example is the BPS string in the rank 1 component of the 9d moduli space obtained as M theory on the M\"{o}bius strip \cite{Aharony:2007du}, where there is accidental supersymmetry enhancement from $(8,0)$ to $(8,8)$ at low energies. While we believe the above is morally correct, we cannot argue that these symmetries must be exact in the worldvolume theory; this is why the previous argument using the geometry of the $U(1)$ instanton moduli space is required.
Furthermore, only in the $S^1/\mathbb{Z}_2$ case non-Abelian symmetries arise \cite{Aharony:2007du}. In the following, we will consider only this case.

%---------------------

The low energy $U(1)$ theory at the general point of the Coulomb branch is specified by the prepotential $\mathcal{F}(\phi)$, which is at most cubic.
The prepotential of 5d $\mathcal{N}=1$ rank $1$ theory is
\begin{align}
\mathcal{F}=\frac{1}{2g^2}\phi^2 + \sum_i\frac{c_i}{6}\bigg( |\phi - \phi_i|^3 + |\phi + \phi_i|^3 \bigg),
\end{align}
where we take $\phi=0$ and $\phi_e$ are the endpoints of the interval $S^1/\mathbb{Z}_2$, and $g, c_i$ are parameters. 
If the endpoint theory is a SCFT, then the gauge coupling is infinite.
The cubic term $c_i$ is only generated by a one-loop computation~\cite{Witten:1996qb}, where the field which becomes light at $\phi=\pm\phi_i$ contributes.
In principle, there could be a tree-level contribution to the cubic term, but it is absent due to the $\mathbb{Z}_2$ quotient.
Note that the effective prepotential on the Coulomb branch is valid even if SCFTs do not admit a gauge theory description. 
The effective gauge coupling is given by the second-order derivative of prepotential:
\begin{align}\label{coupling}
\frac{1}{g^2(\phi)}=\frac{\partial^2 \mathcal{F}}{\partial \phi^2}=\frac{1}{g^2}+\sum_i c_i \bigg(|\phi-\phi_i| + |\phi+\phi_i|\bigg).
\end{align}
The Coulomb branch moduli space metric is
\begin{align}
ds^2 = \frac{1}{g^2(\phi)} d\phi^2.
\end{align}
The Chern-Simons term is given by the third-order derivative of a prepotential:
\begin{align}
\frac{\partial^3 \mathcal{F}}{\partial \phi^3}  \frac{1}{24\pi^2}A\wedge F\wedge F
=\sum_i c_i \,\bigg(\mathrm{sign}\left(\phi-\phi_i\right)+\mathrm{sign}\left(\phi+\phi_i\right)\bigg)\frac{1}{24\pi^2}A\wedge F\wedge F.
\end{align}
Thus, the coefficient $c_i$ represents the ``jump" $\Delta k_i$ in the Chern-Simons term.

The consistency condition can be obtained as follows: consider a double cover $S^1$ with the interval $S^1/\mathbb{Z}_2$.
Let us move the scalar field $\phi$ around $S^1$ once.
At this time, the level of the Chern-Simons term (or the gauge coupling) must come back to its original value.
Since the coefficient $c$ of the cubic term corresponds to the jump in level, the well-definedness of the level, when we come back to the same point, requires that
\begin{align}
\sum_i c_i=\sum_i \Delta k_i=0.
\label{Eq:9d_condition}
\end{align}
This equation, which comes from the compactness of the brane moduli, which in turn comes from the finiteness of black hole entropy, can be used to constrain the bulk gauge symmetry\footnote{This condition can also be used to exclude the existence of non-Abelian symmetries when the Coulomb branch geometry is $S^1$.}.
At any point in the Coulomb branch, the worldvolume theory flows to SCFT or IR free theory at low energy.
Suppose we have a complete classification of 5d rank-1 theories and know the coefficients $c_i$ and the extended global symmetry. 
Then from \eqref{Eq:9d_condition}, we can restrict the possible global symmetries. This translates into a restriction to the bulk gauge symmetry.
Interestingly, as we will soon review, the classification of 5d theory has developed significantly in recent years.
This helps us to obtain new swampland constraints.

In the following, we first review the classification of 5d SCFTs. Then, we list the IR-free theories whose Higgs branch is isomorphic to the instanton moduli space.

\subsubsection*{Classification of 5d SCFTs}
A 5d gauge theory is not renormalizable, because the gauge coupling has a negative mass dimension.
However, if the theory has nontrivial UV fixed points, it can become UV complete as a field theory.
Originally, 5d SCFTs were discovered as theories on $D4$-branes~\cite{Seiberg:1996bd}. 
A large class of 5d SCFTs is obtained by M-theory on local CY threefold with shrinking 4-cycle~\cite{Douglas:1996xp}, and $(p,q)$ 5-brane webs in type IIB string theory~\cite{Aharony:1997bh}.

In recent years, there has been significant progress in the classification of 5d SCFTs~\cite{DelZotto:2017pti,Jefferson:2017ahm,Jefferson:2018irk,Bhardwaj:2018yhy,Bhardwaj:2018vuu,Apruzzi:2018nre,Closset:2018bjz,Apruzzi:2019vpe,Bhardwaj:2019jtr,Apruzzi:2019opn,Apruzzi:2019enx,Bhardwaj:2019fzv,Bhardwaj:2019xeg,Bhardwaj:2020gyu,Bhardwaj:2020kim}.
There are several classification methods. For example, there is a classification based on geometry~\cite{Jefferson:2018irk,Bhardwaj:2019jtr}, a classification based on gauge theory description~\cite{Jefferson:2017ahm,Bhardwaj:2020gyu}, and a classification based on the $S^1$ compactification of the 6d SCFT (which may involve a twist)~\cite{Jefferson:2018irk,Bhardwaj:2018yhy,Bhardwaj:2018vuu,Bhardwaj:2019fzv,Bhardwaj:2020kim}.
In particular, all 5d SCFTs are conjectured to be obtained as RG flows of the 5d KK theory (which is a compactification of 6d SCFT)~\cite{Jefferson:2018irk}.

\begin{figure}[t!]
\begin{center}
    \includegraphics[width=.5\textwidth]{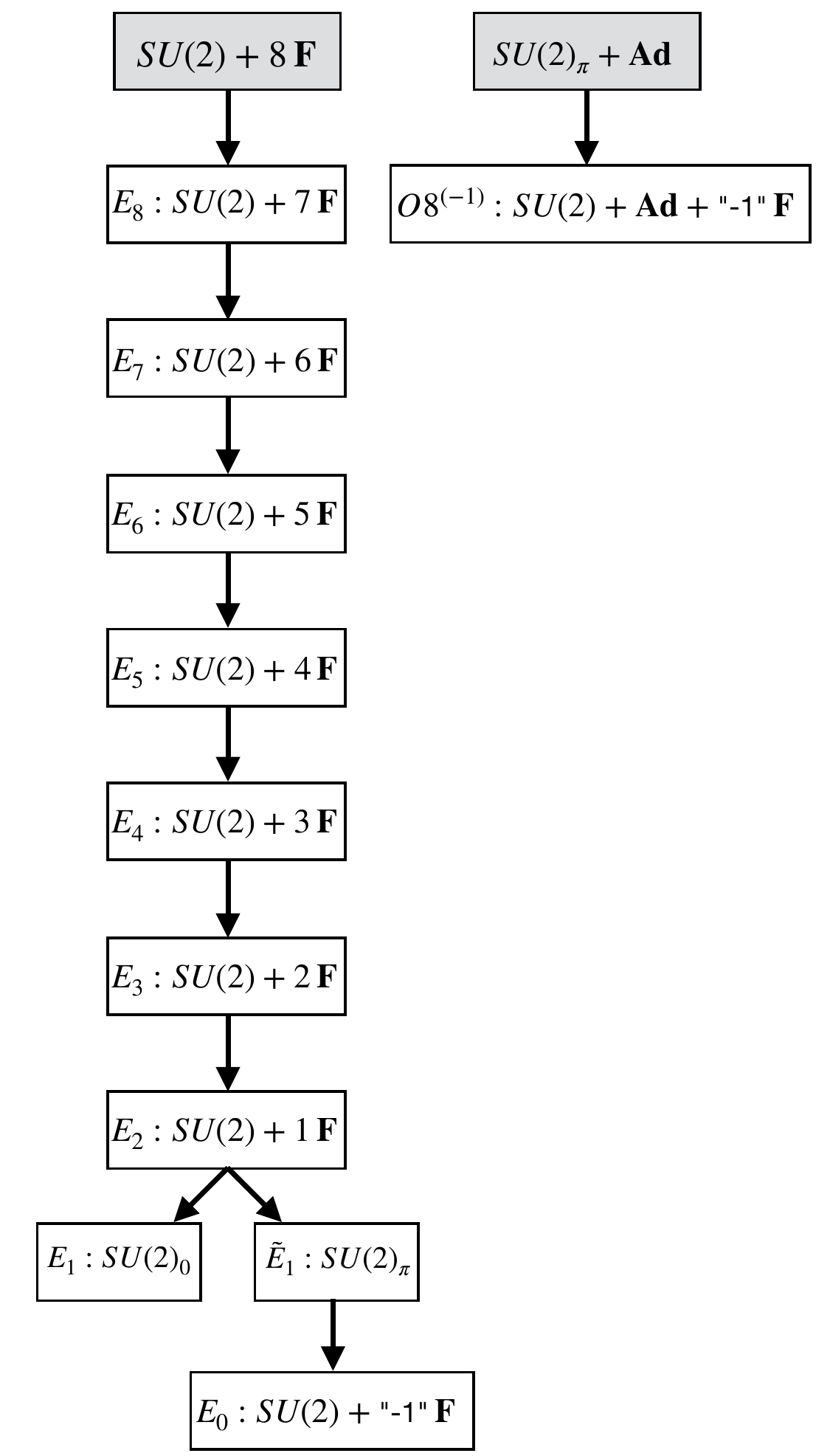}\hfill
\end{center}
\caption{The RG flow among 5d rank 1 SCFTs obtained by mass deformations. 
The shaded boxes correspond to 5d KK theories which are UV completed by 6d SCFTs.
Each box represents a gauge theory description. Since it is rank 1, the gauge group is $SU(2)$ in all cases, where $\mathbf{F}$ represents matter in the fundamental representation and $\mathbf{Ad}$ represents matter in the adjoint representation. In the case of pure gauge theory and adjoint matter only, discrete theta angles are possible, which are denoted by subscripts. There are no gauge theory descriptions for $E_0$ and $O8^{(-1)}$ theories, but we write ``$-1$"$\mathbf{F}$ because RG flow corresponds to formally removing the fundamental representation matter.
}
\label{Fig:5d_SCFTs}
\end{figure}

Here we sketch the classification based on 5d gauge theories obtained from supersymmetry preserving relevant deformations. 
The necessary condition for obtaining nontrivial SCFTs are proposed in \cite{Jefferson:2017ahm}, which is an improved version of \cite{Intriligator:1997pq}.
The conditions are that there exists a physical Coulomb branch where the monopole string has positive tension and the instanton particle has positive mass squared and that the gauge coupling there is positive.
The possible gauge groups of rank 1 are $U(1), O(2)$, and $SU(2)$, of which only $SU(2)$ satisfies the necessary condition.
The $SU(2)$ gauge theories with $N_f=0,1,\cdots,7$ fundamental matters correspond to the gauge coupling deformation of $E_{N_f+1}$ theory, as listed in Fig.~\ref{Fig:5d_SCFTs}.
For $N_f=0$, there exists a freedom to add discrete theta angle~\cite{Douglas:1996xp}.
The SCFT that corresponds to the theory with discrete theta angle $\pi$ is $\tilde{E}_1$ theory.
The $\tilde{E}_1$ theory is amenable to further relevant deformation, which yields the $E_0$ theory.
Note that the $SU(2)$ gauge theory with $N_f=8$ fundamental matters is UV completed by a 6d SCFT.
All $E_n$ and $\tilde{E}_1$ theories are obtained from the RG flow of the 6d SCFT, and the Coulomb branch geometries are locally $\mathbb{R}/\mathbb{Z}_2$. 
The SCFTs are realized at the fixed point.

Similarly, the $SU(2)$ gauge theories with an adjoint matter and discrete theta angle $\pi$ is UV completed by 6d SCFT. 
The relevant deformation of this theory flows to another 5d SCFT in the IR (right side of Fig.~\ref{Fig:5d_SCFTs}).
This theory was found in \cite{Bhardwaj:2019jtr}.
We call this $O8^{(-1)}$ theory because, as we will see below, it is natural to regard this as the worldvolume theory of the $D4$-brane probing the $O8^{(-1)}$-plane.
Here the $O8^{(-1)}$-plane is the orientifold plane whose $D8$-brane charge is $-1$~\cite{Aharony:2007du}.  
The Coulomb branch geometry of the $O8^{(-1)}$ theory is $\mathbb{R}/\mathbb{Z}_2$.

If we include the matter in the $SU(2)$ gauge theory with representations that do not appear in Fig.~\ref{Fig:5d_SCFTs}, the value of $c_i$ becomes negative. This makes the gauge coupling negative and does not satisfy the necessary condition in \cite{Jefferson:2017ahm}.

In this paper, we derive a consistency condition of the worldvolume theory of probe branes, assuming that the classification above is complete.
In principle, it is possible that there exist unknown SCFTs which do not admit a gauge theory description and are associated with unknown geometry, although we believe that this is unlikely.

The values of $c_i$ in the SCFTs described above are as follows:~\cite{Seiberg:1996bd,Morrison:1996xf,Bhardwaj:2019jtr}
\begin{align}
\frac{c}{c_{A_0}}= \begin{cases}
9-n \quad \text{for $E_n$  theory ($n=0,1,\cdots,8$)}\\
8 \quad \text{for $\tilde{E}_1$  theory}\\
1 \quad \text{for $O8^{(-1)}$  theory}
\end{cases},
\label{Eq:c_values}\end{align}
where $c_{A_0}$ is the value of $c$ for $A_0$ theory ($U(1)$ gauge theory with one electron).

\subsubsection*{IR free theories}
The theory of the symmetry enhanced point on the 4-brane can be an IR-free theory as well as a SCFT. 
Here we list the rank-1 free theories in which the Higgs branche is isomorphic to an instanton moduli space.

There are two theories in which the Higgs branch is a one-instanton moduli space: $A_n$ theory and $D_n$ theory.
The $A_n$ theory is a $U(1)$ gauge theory containing $(n+1)$ ``electrons" as matter. This theory has an $\mathfrak{su}(n+1)$ global symmetry.
A $D_n$ theory is a $SU(2)$ gauge theory containing $n$ ``quarks" as matter. This theory has an $\mathfrak{spin}(2n)$ global symmetry.
The Coulomb branch geometry of $A_n$ theory is $\mathbb{R}$ and that of $D_n$ theory is $\mathbb{R}/\mathbb{Z}_2$ considering the Weyl group.
Compactifying the $A_n$ theory to $S^1$ results in a 4d $\mathcal{N}=2$ theory corresponding to the $\mathfrak{su}(n+1)$ small instanton in 8d spacetime.
This corresponds to the $I_n$ singularity of the 4d Coulomb branch.
Similarly, the $S^1$ compactification of the $D_n$ theory is the $I_n^*$ singularity of the 4d Coulomb branch.

For the $C_n$ case, the one-instanton Higgs branch is simply given by a free half-hypermultiplet, and has a zero-dimensional Coulomb branch. This means that the corresponding probe brane is ``stuck'', and indeed this theory is realized by the worldvolume of a stuck $D4$ branes. By contrast, the theory with instanton number of two has a one-dimensional Coulomb branch, and as argued in \cite{Hamada:2021bbz} it is the one describing mobile probe branes that connect to small instantons of other non-Abelian factors. Therefore, we will focus on this theory, which is given by an $O(2)=U(1)\rtimes\mathbb{Z}_2$ gauge theory containing two hypermultiplets of charge $2$ and $\mathfrak{sp}(n)$ hypermultiplets of fundamental representation as matter~\cite{Witten:1997bs}. 
The global symmetry of this theory is $\mathfrak{sp}(n)$.
The Coulomb branch geometry of $C_n$ theory is $\mathbb{R}/\mathbb{Z}_2$ because of a discrete gauging.
The $S^1$ compactification of this theory is the frozen $I_{n+8}^*$ singularity in the 4d Coulomb branch.
The lack of mass deformation due to $\mathbb{Z}_2$ gauging in the $O(2)$ gauge symmetry corresponds to the singularity being frozen~\cite{Witten:1997bs}.

The jump in the level of the Chern-Simons term (which is the same as the change in the slope of the gauge coupling) can be obtained exactly by a 1-loop calculation~\cite{Witten:1996qb}. The result is 
\begin{align}
\frac{c}{c_{A_0}}= \begin{cases}
-\left(n+1\right) \quad \text{for $A_n$ theory}\\
8-n \quad \text{for $D_n$  theory}\\
-(8+n)  \quad \text{for $C_n$  theory}
\end{cases}.
\label{Eq:c_values_IR_free}\end{align}

One could also ask why these are all the possibilities for the global symmetry of IR free theories. Naively, these are easy to construct: For any $G$, just consider $U(1)$ gauge theory with matter in a representation of $G$. At low energies, the scalars in the matter sector are described just by a kinetic term
\begin{equation} \mathcal{L}\supset \int \sqrt{-g}\left[\frac{\kappa_{ab}}{2}\partial_\mu \phi^a\partial^\mu\phi^b\right],\label{45rr}\end{equation}
where the indices $a,b$ take values in some representation of $G$, and $\kappa_{ab}$ is the corresponding quadratic form.
At low energies, this theory has a $G$ global symmetry. In particular, there seems to be no obstacle to things like $G=G_2$ in eight dimensions, which we know does not arise in the landscape of known 8d $\mathcal{N}=1$ theories \cite{Hamada:2021bbz}. 

However, for any theory with lagrangian \eqref{45rr}, the global symmetry in the IR enhances to $Spin(n)$,  $Sp(n)$, or $U(n)$, according to whether the representation under consideration is real, pseudoreal, or complex, respectively. The point is that any finite-dimensional representation of any group comes with the quadratic form $\kappa_{ab}$ that one uses to construct the non-degenerate kinetic term, and that the symmetry of the lagrangian \eqref{45rr} is that of the quadratic form. When the representation is real, $\kappa_{ab}$ is a real symmetric matrix, and the corresponding symmetry group is the orthogonal group. When the representation is pseudoreal, it preserves a symplectic form; the symmetry is the symplectic group. And when the representation is complex, $\kappa_{ab}$ preserves a Hermitian form, whose symmetry group is unitary.

So, assuming there is no accidental symmetry enhancement in the deep IR, the only possibilities are the ones we have listed. We will now use these to classify the possible consistent quantum gravity vacua that one can have in nine dimensions. 

\subsubsection*{Comparison of \eqref{Eq:9d_condition} and string vacua}
\begin{table}[!t]
  \renewcommand{\arraystretch}{1.5}
    \addtolength{\tabcolsep}{1pt} 
\begin{center}
\small
\begin{tabular}{|c|ccccc|}\hline \hline
  Name    &  free or CFT& Symmetry &  Geometry      & Brane & $c/c_{A_0}$          \\\hline\hline
$A_n (n=0,\cdots)$   & free & $\mathfrak{su}(n+1)$   &  $\mathbb{R}$  & $(n+1)D8$ &$-(n+1)$  \\\hline
$C_n (n=0,\cdots)$   & free & $\mathfrak{sp}(n)$  &  $\mathbb{R}/\mathbb{Z}_2$   & $O8^++nD8$ & $-(8+n)$   \\\hline
$D_n (n=0,\cdots)$   & free & $\mathfrak{spin}(2n)$ &  $\mathbb{R}/\mathbb{Z}_2$  & $O8^-+nD8$ & $8-n$  \\\hline
$E_n (n=1,\cdots,8)$   & CFT & caption  &  $\mathbb{R}/\mathbb{Z}_2$   & $O8^-+(n-1)D8$ & $9-n$ \\\hline
$\tilde{E}_1 $   & CFT & $\mathfrak{u}(1)$  &  $\mathbb{R}/\mathbb{Z}_2$   & $O8^-$ & $8$ \\\hline
$E_0 $   & CFT & $\varnothing$  &  $\mathbb{R}/\mathbb{Z}_2$   & $O8^{(-9)}$ & $9$ \\\hline
$O8^{(-1)}$   & CFT & $\varnothing$  &  $\mathbb{R}/\mathbb{Z}_2$   & $O8^{(-1)}$ & $1$ \\\hline
\end{tabular}
\caption{List of 4-brane worldvolume theories with one-dimensional Coulomb branch.
Global symmetries of $E_n$ and $\tilde{E}_1$ theories are $E_8=\mathfrak{e}_8, E_7=\mathfrak{e}_7, E_6=\mathfrak{e}_6, E_5=\mathfrak{spin}(10), E_4=\mathfrak{su}(5), E_3=\mathfrak{su}(3)+\mathfrak{su}(2), E_2=\mathfrak{su}(2)+\mathfrak{u}(1), E_1=\mathfrak{su}(2), \tilde{E}_1=\mathfrak{u}(1), E_0=\varnothing$.
The Higgs branch is given by one-instanton moduli space (two-instantons moduli space for $C_n$).
The geometry column refers to the local geometry around the symmetry enhanced point.
See Eqs.~\eqref{Eq:c_values} and \eqref{Eq:c_values_IR_free} for the details of the $c/c_{A_0}$ column, and the text after \eqref{Eq:9d_string_vacua} for details behind the brane column.
}
\label{Tab:4-brane_theories}
\end{center}
\end{table}

So far, we have listed the possible brane worldvolume theories that can arise at the endpoints or singular points of the  $S^1/\mathbb{Z}_2$ Coulomb branch moduli space. These are summarized in Table~\ref{Tab:4-brane_theories}.
Since the overall geometry is $S^1/\mathbb{Z}_2$, there are worldvolume theories with a geometry of $\mathbb{R}/\mathbb{Z}_2$ at the two endpoints.
It is important to emphasize that there should be exactly two worldvolume theories with $\mathbb{R}/\mathbb{Z}_2$ geometries. 

Only the $A_n$ theory can appear as a singularity inside a line segment.
The fact that the cubic coefficient $c_i$ in $A_n$ theory is negative is important in obtaining the bound.

The different choices of theories on the two endpoints correspond to different classes of vacua.
There are 10 ways to choose two endpoints from $C_n, D_n, E_n$ (including $\tilde{E}_1$), and $O8^{(-1)}$ theories. 
However, in order to satisfy \eqref{Eq:9d_condition}, given that $c_{A_n}$ is negative, the sum of $c_i$ of the theory on the endpoints must be non-negative.
This excludes the case where both endpoint theories are $C_n$ and the case where $C_n$ and $O8^{(-1)}$ are chosen.
The remaining eight patterns are as follows:
\begin{align}
&2\,D, D+E, 2\,E: \text{Rank $17$ theories}, \nonumber \\
&D+O8^{(-1)}, E+O8^{(-1)}: \text{Rank $9$ theories}, \nonumber \\
&2\,O8^{(-1)}: \text{Rank $1$ theories A}, \nonumber \\
&C+D, C+E: \text{Rank $1$ theories B}.
\label{Eq:probe_patterns}
\end{align}
Here, on the right side, we have written the rank of the symmetry group, which is consistent with the result in  \cite{Montero:2020icj}.

The non-Abelian symmetry can be read from the global symmetry groups at the singularity, but it is not a priori clear how to count the number of $\mathfrak{u}(1)$ symmetries. Here we make two important comments about the counting of $\mathfrak{u}(1)$ factors. Later in the Subsection, we will provide a more complete and systematic study of the $\mathfrak{u}(1)$ factors.

First, there are $\mathfrak{u}(1)$'s associated with the relative positions of $A_n$ singularities.
This is understood by starting from the $A_1$ singularity and deforming it.
The 5d theory at the $A_1$ singularity has an $\mathfrak{su}(2)$ global symmetry that rotates the two electrons, which corresponds to the bulk gauge symmetry.
Then, let us consider breaking the $\mathfrak{su}(2)$ symmetry. This is achieved, in the bulk, by giving the vacuum expectation value to the Cartan component of the scalar field in the vector multiplet. On the brane, on the other hand, it is achieved by giving a mass difference to the two electrons (corresponding to an $A_1$ singularity splits into two $A_0$ singularities).
Therefore, we see that the $\mathfrak{u}(1)$ vector multiplet in the bulk is coupled to the mass difference operator on the brane.
Similar arguments show that the relative position of $A_n$ and $D_m$ singularities are also associated with $\mathfrak{u}(1)$'s.

Next, there may be an additional $\mathfrak{u}(1)$ corresponding to the 5d instanton number.
If a 5d theory at the fixed point is IR free, then there is a conserved current \cite{Seiberg:1996bd}
\begin{align}
j = * \,\mathrm{Tr}\left(F\wedge F\right).
\end{align}
This generates an $\mathfrak{u}(1)$ symmetry under which the BPS instanton particle is charged.
As before, this must couple to the 9d bulk vector multiplet in a supersymmetric way.
This means that there is a $\mathfrak{u}(1)$ gauge symmetry in the bulk and that the 5d gauge coupling (and mass of instanton particle) is controlled by the corresponding 9d scalar field.

We will now explain how the gauge group (including its abelian factor) can be completely determined from the data of the Coulomb branch of the brane theory. The data that we are interested in is the collection of global symmetries of the brane worldvolume theory and the points where those global symmetries are realized on the Coulomb branch. But for now, we consider theories where the relative positions of all points of symmetry enhancement on the Coulomb branch are frozen. The positions can freeze due to the continuity of the brane coupling constant $g$ across the brane moduli space. For example, consider the case where there are two $E_8$ theories at the endpoints and an $A_0$ theory somewhere in the middle. In such a theory, the location of the $A_0$ is forced to be exactly at the center of the interval. This is because $1/g^2$ vanishes at the $E_8$ endpoints and symmetrically increases in the middle. Therefore, the tipping point of $1/g^2$,  where $A_0$ is located, must be at the center. 

In theories where the relative positions of points of symmetry enhancement for a fixed gauge group are completely frozen, enhancing the symmetry algebra is impossible\footnote{Note that our definition for maximally enhanced theory is one where the gauge group cannot be further enhanced to a larger group. This is different from the definition in \cite{Font:2020rsk} where maximal enhancement refers to the absence of $\mathfrak{u}(1)$ factors.}. Later we will show that these theories are the only maximally enhanced theories. In other words, we show that the gauge group can always be enhanced to one of these symmetry groups.

In most cases, the maximally enhanced theory has a semisimple symmetry algebra of rank of 17, 9, or 1. However, in a few cases, the rank of the semisimple algebra is off by one. Therefore, for such theories to be realized, there must be an extra $\mathfrak{u}(1)$ in the gauge group. All the frozen geometries in the sense discussed above and their corresponding gauge algebras are listed in the following tables. The theories with ranks 17, 9, and 1, are respectively listed in Tables~\ref{Tab:rank17}, \ref{Tab:rank9}, and \ref{Tab:rank1}.

\begin{center}
\small
  \renewcommand{\arraystretch}{1.5}
    \addtolength{\tabcolsep}{1pt} 
\begin{longtable}{|c|c|c|c|}\hline \hline
  \# & \makecell{Placement of enhanced \\theories on the Coulumb branch}    &  Gauge algebra & Root lattice          \\\hline\hline
1 & $E_8\hdash A_1 \hdash E_8$   & $\mathfrak{e}_8+\mathfrak{e}_8+\mathfrak{su}(2)$ & $2E_8+A_1$\\\hline
2 & $E_8\hdash A_2 \hdash E_7$   & $\mathfrak{e}_8+\mathfrak{e}_7+\mathfrak{su}(3)$ & $E_8+E_7+A_2$\\\hline
3 & $E_8\hdash A_3 \hdash E_6$   & $\mathfrak{e}_8+\mathfrak{e}_6+\mathfrak{su}(4)$ & $E_8+E_6+A_3$\\\hline
4 & $E_8\hdash A_4 \hdash E_5$   & $\mathfrak{e}_8+\mathfrak{spin}(10)+\mathfrak{su}(5)$ & $E_8+D_5+A_4$\\\hline
5 & $E_8\hdash A_5 \hdash E_4$   & $\mathfrak{e}_8+\mathfrak{su}(6)+\mathfrak{su}(5)$ & $E_8+A_5+A_4$\\\hline
6 & $E_8\hdash A_6 \hdash E_3$   & $\mathfrak{e}_8+\mathfrak{su}(7)+\mathfrak{su}(3)+\mathfrak{su}(2)$ & $E_8+A_6+A_2+A_1$\\\hline
7 & $E_8\hdash A_8 \hdash E_1$   & $\mathfrak{e}_8+\mathfrak{su}(9)+\mathfrak{su}(2)$ & $E_8+A_8+A_1$\\\hline
8 & $E_8\hdash A_9 \hdash E_0$   & $\mathfrak{e}_8+\mathfrak{su}(10)$ & $E_8+A_9$\\\hline
9 & $E_7\hdash A_3 \hdash E_7$   & $\mathfrak{e}_7+\mathfrak{e}_7+\mathfrak{su}(4)$ & $2E_7+A_3$\\\hline
10 & $E_7\hdash A_4 \hdash E_6$   & $\mathfrak{e}_7+\mathfrak{e}_6+\mathfrak{su}(5)$ & $E_7+E_6+A_4$\\\hline
11 & $E_7\hdash A_5 \hdash E_5$   & $\mathfrak{e}_7+\mathfrak{spin}(10)+\mathfrak{su}(6)$ & $E_7+D_5+A_5$\\\hline
12 & $E_7\hdash A_6 \hdash E_4$   & $\mathfrak{e}_7+\mathfrak{su}(7)+\mathfrak{su}(5)$ & $E_7+A_6+A_4$\\\hline
13 & $E_7\hdash A_7 \hdash E_3$   & $\mathfrak{e}_7+\mathfrak{su}(8)+\mathfrak{su}(3)+\mathfrak{su}(2)$ & $E_7+A_7+A_2+A_1$\\\hline
14 & $E_7\hdash A_9 \hdash E_1$   & $\mathfrak{e}_7+\mathfrak{su}(10)+\mathfrak{su}(2)$ & $E_7+A_9+A_1$\\\hline
15 & $E_7\hdash A_{10} \hdash E_0$   & $\mathfrak{e}_7+\mathfrak{su}(11)$ & $E_7+A_{10}$ \\\hline
16 & $E_6\hdash A_5 \hdash E_6$   & $\mathfrak{e}_6+\mathfrak{e}_6+\mathfrak{su}(6)$ & $2E_6+A_5$\\\hline
17 & $E_6\hdash A_6 \hdash E_5$   & $\mathfrak{e}_6+\mathfrak{spin}(10)+\mathfrak{su}(7)$ & $E_6+D_5+A_6$\\\hline
18 & $E_6\hdash A_7 \hdash E_4$   & $\mathfrak{e}_6+\mathfrak{su}(8)+\mathfrak{su}(5)$ & $E_6+A_7+A_4$\\\hline
19 & $E_6\hdash A_8 \hdash E_3$   & $\mathfrak{e}_6+\mathfrak{su}(9)+\mathfrak{su}(3)+\mathfrak{su}(2)$ & $E_6+A_8+A_2+A_1$\\\hline
20 & $E_6\hdash A_{10} \hdash E_1$   & $\mathfrak{e}_6+\mathfrak{su}(11)+\mathfrak{su}(2)$ & $E_6+A_{10}+A_1$\\\hline
21 & $E_6\hdash A_{11} \hdash E_0$   & 
$\mathfrak{e}_6+\mathfrak{su}(12)$ & $E_6+A_{11}$\\\hline
22 & $E_5\hdash A_7 \hdash E_5$   & $\mathfrak{spin}(10)+\mathfrak{spin}(10)+\mathfrak{su}(8)$ & $2D_5+A_7$\\\hline
23 & $E_5\hdash A_8 \hdash E_4$   & $\mathfrak{spin}(10)+\mathfrak{su}(9)+\mathfrak{su}(5)$ & $D_5+A_8+A_4$\\\hline
24 & $E_5\hdash A_9 \hdash E_3$   & \makecell{$\mathfrak{spin}(10)+\mathfrak{su}(10)$\\$+\mathfrak{su}(3)+\mathfrak{su}(2)$} & $D_5+A_9+A_2+A_1$\\\hline
25 & $E_5\hdash A_{11} \hdash E_1$   & $\mathfrak{spin}(10)+\mathfrak{su}(12)+\mathfrak{su}(2)$ & $D_5+A_{11}+A_1$\\\hline
26 & $E_5\hdash A_{12} \hdash E_0$   & 
$\mathfrak{spin}(10)+\mathfrak{su}(13)$ & $D_5+A_{12}$\\\hline
27 & $E_4\hdash A_9 \hdash E_4$   & $\mathfrak{su}(10)+\mathfrak{su}(5)+\mathfrak{su}(5)$ & $A_9+2A_4$\\\hline
28 & $E_4\hdash A_{10} \hdash E_3$   & \makecell{$\mathfrak{su}(11)+\mathfrak{su}(5)$\\$+\mathfrak{su}(3)+\mathfrak{su}(2)$} & $A_{10}+A_4+A_2+A_1$\\\hline
29 & $E_4\hdash A_{12} \hdash E_1$   & $\mathfrak{su}(13)+\mathfrak{su}(5)+\mathfrak{su}(2)$ & $A_{12}+A_4+A_1$\\\hline
30 & $E_4\hdash A_{13} \hdash E_0$   & 
$\mathfrak{su}(14)+\mathfrak{su}(5)$ & $A_{13}+A_4$\\\hline
31 & $E_3\hdash A_{11} \hdash E_3$   & \makecell{$\mathfrak{su}(12)+\mathfrak{su}(3)+\mathfrak{su}(3)$\\$+\mathfrak{su}(2)+\mathfrak{su}(2)$} & $A_{11}+2A_2+2A_1$\\\hline
32 & $E_3\hdash A_{13} \hdash E_1$   & \makecell{$\mathfrak{su}(14)+\mathfrak{su}(3)$\\$+\mathfrak{su}(2)+\mathfrak{su}(2)$ }& $A_{13}+A_2+2A_1$\\\hline
33 & $E_3\hdash A_{14} \hdash E_0$   & 
$\mathfrak{su}(15)+\mathfrak{su}(3)+\mathfrak{su}(2)$ & $A_{14}+A_2+A_1$\\\hline
34 & $E_1\hdash A_{15} \hdash E_1$   & $\mathfrak{su}(16)+\mathfrak{su}(2)+\mathfrak{su}(2)$ & $A_{15}+2A_1$ \\\hline
35 & $E_1\hdash A_{16} \hdash E_0$   & 
$\mathfrak{su}(17)+\mathfrak{su}(2)$ & $A_{16}+A_1$\\\hline
36 & $E_0\hdash A_{17} \hdash E_0$   & 
$\mathfrak{su}(18)$ & $A_{17}$\\\hline
37 & $E_8\fdash D_9$ & $\mathfrak{e}_8+\mathfrak{spin}(18)$ & $E_8+D_9$\\\hline
38 & $E_7\fdash D_{10}$ & $\mathfrak{e}_7+\mathfrak{spin}(20)$ & $E_7+D_{10}$\\\hline
39 & $E_6\fdash D_{11}$ & $\mathfrak{e}_6+\mathfrak{spin}(22)$ & $E_6+D_{11}$\\\hline
40 & $E_5\fdash D_{12}$ & $\mathfrak{spin}(24)+\mathfrak{spin}(10)$ & $D_{12}+D_5$\\\hline
41 & $E_4\fdash D_{13}$ & $\mathfrak{su}(5)+\mathfrak{spin}(26)$ & $D_{13}+A_4$\\\hline
42 & $E_3\fdash D_{14}$ & $\mathfrak{su}(3)+\mathfrak{su}(2)+\mathfrak{spin}(28)$ & $D_{14}+A_2+A_1$\\\hline
43 & $E_1\fdash D_{16}$ & $\mathfrak{su}(2)+\mathfrak{spin}(32)$ & $D_{16}+A_1$\\\hline
44 & $E_0\fdash D_{17}$ & $\mathfrak{spin}(34)$ & $D_{17}$\\\hline
\makecell{45\\-53} & \makecell{$0\leq n\leq8: D_n\fdash D_{16-n}$} & \makecell{$\mathfrak{spin}(2n)+\mathfrak{spin}(32-2n)$\\$+\mathfrak{u}(1)$} & $D_{n}+D_{16-n}$\\\hline
\caption{List of possible maximally enhanced rank 17 theories in nine dimensions which is obtained by Swampland considerations. The first 44 lines where the algebra is semisimple match with the Table 3 in \cite{Font:2020rsk} which have string theory realizations.}
\label{Tab:rank17}
\end{longtable}
\end{center}

\begin{center}
\small
  \renewcommand{\arraystretch}{1.5}
    \addtolength{\tabcolsep}{1pt} 
\begin{longtable}{|c|c|c|c|}\hline \hline
  \# & \makecell{Placement of enhanced \\theories on the Coulumb branch}    &  Gauge algebra & Root lattice          \\\hline\hline
1 & $E_8\hdash A_1 \hdash O8^{(-1)}$   & $\mathfrak{e}_8+\mathfrak{su}(2)$ & $E_8+A_1$\\\hline
2 & $E_7\hdash A_2 \hdash O8^{(-1)}$   & $\mathfrak{e}_7+\mathfrak{su}(3)$ & $E_7+A_2$\\\hline
3 & $E_6\hdash A_3 \hdash O8^{(-1)}$   & $\mathfrak{e}_6+\mathfrak{su}(4)$ & $E_6+A_3$\\\hline
4 & $E_5\hdash A_4 \hdash O8^{(-1)}$   & $\mathfrak{spin}(10)+\mathfrak{su}(5)$ & $D_5+A_4$\\\hline
5 & $E_4\hdash A_5 \hdash O8^{(-1)}$   & $\mathfrak{su}(6)+\mathfrak{su}(5)$ & $A_5+A_4$\\\hline
6 & $E_3\hdash A_6 \hdash O8^{(-1)}$   & $\mathfrak{su}(7)+\mathfrak{su}(3)+\mathfrak{su}(2)$ & $A_6+A_2+A_1$\\\hline
7 & $E_1\hdash A_8 \hdash O8^{(-1)}$   & $\mathfrak{su}(9)+\mathfrak{su}(2)$ & $A_8+A_1$\\\hline
8 & $E_0\hdash A_9 \hdash O8^{(-1)}$   & $\mathfrak{su}(10)$ & $A_9$\\\hline
9 & $D_9\fdash O8^{(-1)}$   & $\mathfrak{spin}(18)$ & $D_9$\\\hline
\caption{List of possible maximally enhanced rank 9 theories in nine dimensions which is obtained by Swampland considerations. The above table match with the Table 3 in \cite{Font:2021uyw} which have string theory realizations.}
\label{Tab:rank9}
\end{longtable}
\end{center}

\begin{center}
\small
  \renewcommand{\arraystretch}{1.5}
    \addtolength{\tabcolsep}{1pt} 
\begin{longtable}{|c|c|c|c|}\hline \hline
  \# & \makecell{Placement of enhanced \\theories on the Coulumb branch}    &  Gauge algebra & Root lattice          \\\hline\hline
1 & $O8^{(-1)}\hdash A_1 \hdash O8^{(-1)}$   & $\mathfrak{su}(2)$ & $A_1$\\\hline
2 & $E_1\fdash C_0$   & $\mathfrak{su}(2)$ & $A_1$\\\hline
3 & $E_0\fdash C_1$   & $\mathfrak{su}(2)$ & $A_1$\\\hline
4 & $D_0\fdash C_0$   & $\mathfrak{u}(1)$ & $\varnothing$\\\hline
\caption{List of possible maximally enhanced rank 1 theories in nine dimensions which is obtained by Swampland considerations.}
\label{Tab:rank1}
\end{longtable}
\end{center}

Let us compare the above with string compactifications. 
In string theory, there are four classes of vacua~\cite{deBoer:2001wca,Keurentjes:2001cp,Aharony:2007du,Kim:2019ths,Font:2020rsk,Font:2021uyw}.
To make the comparison with \eqref{Eq:probe_patterns} easier to understand, we write each class as follows.
\begin{align}
&2\,O8^- + 16 D8: \text{Rank $17$ theories},\nonumber\\
&O8^- + O8^0 + 8 D8: \text{Rank $9$ theories},\nonumber\\
&2\,O8^0: \text{Rank $1$ theories A}, \nonumber\\
&O8^- + O8^+: \text{Rank $1$ theories B},
\label{Eq:9d_string_vacua}\end{align}
where $O8^0$ is the shift orientifold~\cite{Keurentjes:2001cp,Aharony:2007du}, and $O8^{\pm}$ are the orientifold planes with D8-charge $\pm8$.

Theories with rank 17 are obtained by circle compactification of heterotic/type $I$ strings, as well as type $I'$ string where there are two $O8^-$-planes and 16 $D8$-branes~\cite{Polchinski:1995df}.
In this context, the 4-brane obtained as a small instanton is a $D4$-brane.
The theory on the $D4$-brane that probes the $(n+1)\,D8$-branes is the $A_n$ theory, and the theory on the $D4$-brane that probes the $O8^-+n D8$ is the $D_n$ theory.
When the dilaton diverges at the position of $O8^-+(n-1)\,D8$, then the $E_n$ theory is realized.
It is also possible for $O8^-$ to emit $D8$ non-perturbatively ($O8^-\to O8^{(-9)}+D8$)\footnote{The superscript corresponds to $D8$-brane charge.}.  
The worldvolume theory of a $D4$-brane that probes the $O8^{(-9)}$ becomes the $E_0$ theory.
This is not captured by the perturbative type $I'$ description, but it can be understood from the language of geometry in real K3~\cite{Cachazo:2000ey}.
Table 11 of \cite{Font:2020rsk} lists all the patterns of the maximally enhanced gauge groups in the 9-dimensional heterotic string vacuum, which matches the ones in Table~\ref{Tab:rank17}.

Theories with rank 9 are obtained from the CHL string~\cite{Chaudhuri:1995fk,Chaudhuri:1995bf}, M-theory on the M\"{o}bius strip~\cite{Park:1996it,Dabholkar:1996pc}, and IIA string with $O8^-+O8^0+8D8$~\cite{Aharony:2007du}.
Again, when the $O$-plane emits $D8$ non-perturbatively ($O8^-\to O8^{(-9)}+D8$ and $O8^0\to O8^{(-1)}+D8$), the maximally enhanced gauge symmetries are realized.
By comparing with \eqref{Eq:probe_patterns}, we can see that the $O8^{(-1)}$ theory in Table~\ref{Tab:4-brane_theories} is naturally interpreted as a worldvolume theory for $D4$-brane probing the $O8^{(-1)}$ plane.
Table 3 of \cite{Font:2021uyw} lists all the patterns of the maximally enhanced gauge groups in the 9-dimensional CHL string vacuum, which matches the ones in Table~\ref{Tab:rank9}.

Finally, there are two inequivalent theories that have rank-1.
One is M-theory on the Klein Bottle~\cite{Dabholkar:1996pc}, the Asymmetric Orbifold of IIA, and is IIA with $2\,O8^0$~\cite{Aharony:2007du}.
The other is IIB on the Dabholkar-Park background~\cite{Dabholkar:1996pc}, the Asymmetric Orbifold of IIB, and IIA with $O8^-+O8^+$~\cite{Aharony:2007du}.
It is known that in both classes of theories, the symmetry can be enhanced to $SU(2)$, which matches the list in Table~\ref{Tab:rank1}.

\subsubsection*{Spacetime gauge theory and instanton moduli space}

For brane moduli spaces listed in Tables \ref{Tab:rank17}, \ref{Tab:rank9}, and \ref{Tab:rank1}, we found the corresponding spacetime gauge group and showed that the gauge group is maximally enhanced. In the following, we complete our analysis by determining the spacetime gauge group for the ones not listed in the tables. The semisimple part of the gauge algebra is easy to find as it is given by the global symmetries of the brane theory. However, counting the number of additional $\mathfrak{u}(1)$ components turns out to be non-trivial. Moreover, we will show that the theories listed in the tables are the only maximally enhanced theories. In other words, the gauge group of any theory with a different brane moduli space could be enhanced to one of the entries of Tables \ref{Tab:rank17}, \ref{Tab:rank9}, or \ref{Tab:rank1}.

To show any other theory can be a=enhanced, we look at the deformations of the Coulomb branch of the brane resulting from moving around in the Coulomb branch of the bulk theory. In addition to the continuous change in the position of  $A_n$ points in the interior of the interval, multiple groups can fuse or break up. These can be most easily read off from the string theory realization of these theories. Note that this is a field theory statement, even though we use the string theory realizations of these theories to verify it.  We find that the following transitions are allowed:
\begin{enumerate}\label{transitions}
\item $E_0$ and $A_0$ $\leftrightarrow$ $\tilde E_1$ corresponding to $O8^{(-9)} + D8$ $\leftrightarrow$ $O8^{-}$.
\item $E_1$ and $A_0$ $\leftrightarrow$ $E_2$ corresponding to moving away a $D8$ from the $E_2$ point.
\item $A_m$ and $A_n$ $\leftrightarrow$ $A_{m+n+1}$ corresponding to joining/separating two stack of $m+1$ and $n+1$ $D8$ branes .
\item $A_m$ and $D_n$ $\leftrightarrow$ $D_{m+n+1}$ corresponding to moving to/away a stack of $m+1$ $D8$ branes to/from a stack of $O8^-$ and $n$ $D8$ branes.
\item $C_0$ and $A_0$ $\leftrightarrow$ $C_1$ corresponding to moving joining/separating a $D8$ brane to/from the $O8^+$ brane. 
\end{enumerate}
We did not include transitions involving $C_{n>1}$ since, as we will see later, including such theories makes it impossible to satisfy the condition \eqref{Eq:9d_condition}.

Note that the first two transitions do not change the rank of the semisimple Lie algebra of the gauge group. However, the last three transitions change the rank of the semisimple Lie algebra by one. Since the the total rank of the group is invariant, these transitions must also involve the appearance of additional $\mathfrak{u}(1)$. In other words, the rank change comes from (un)Higgsing mechanism that absorbs/breaks up a $\mathfrak{u}(1)$ to/from the Lie algebra\footnote{This is consistent with the argument below \eqref{Eq:probe_patterns}.}.

Following, we implement an algorithmic series of these transitions that will maximally enhance the gauge group. 

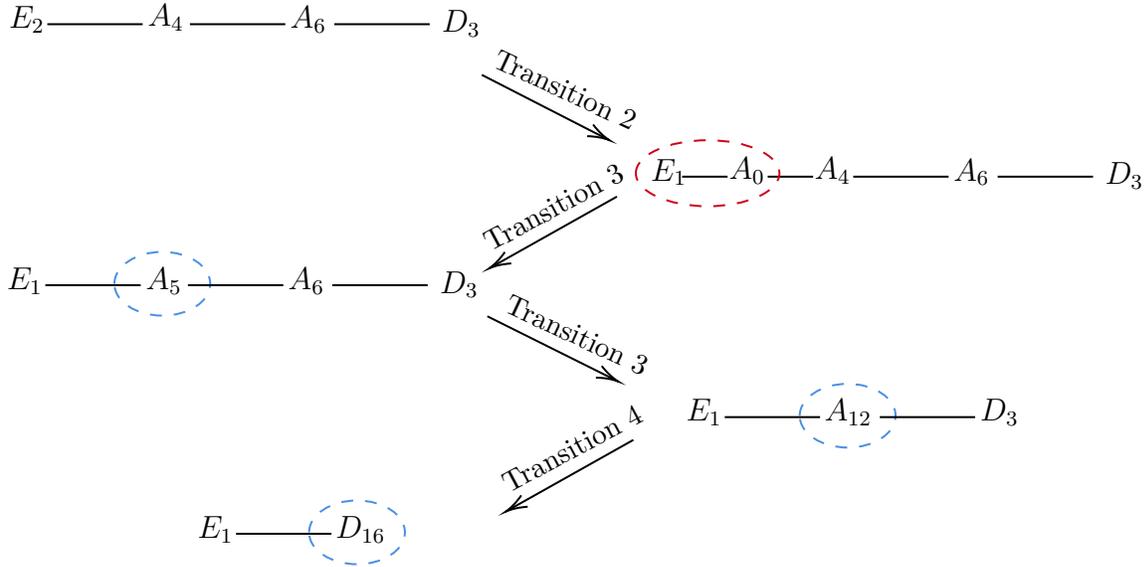
\begin{figure}
    \centering

\tikzset{every picture/.style={line width=0.75pt}} %set default line width to 0.75pt        

\begin{tikzpicture}[x=0.75pt,y=0.75pt,yscale=-1,xscale=1]
%uncomment if require: \path (0,348); %set diagram left start at 0, and has height of 348

%Straight Lines [id:da9946152311454677] 
\draw    (62.68,35.44) -- (110.78,35.44) ;
%Straight Lines [id:da02702443157275769] 
\draw    (134.6,35.44) -- (182.7,35.44) ;
%Straight Lines [id:da6566698913150797] 
\draw    (207.43,35.44) -- (255.53,35.44) ;
%Straight Lines [id:da5282595805645975] 
\draw    (281.82,61) -- (344.4,93.09) ;
\draw [shift={(346.18,94)}, rotate = 207.14] [color={rgb, 255:red, 0; green, 0; blue, 0 }  ][line width=0.75]    (10.93,-3.29) .. controls (6.95,-1.4) and (3.31,-0.3) .. (0,0) .. controls (3.31,0.3) and (6.95,1.4) .. (10.93,3.29)   ;
%Straight Lines [id:da6562408918237386] 
\draw    (426,112.4) -- (448.93,112.4) ;
%Straight Lines [id:da5125727158037643] 
\draw    (469.15,112.4) -- (517.25,112.4) ;
%Straight Lines [id:da7057679371895877] 
\draw    (541.98,112.4) -- (590.08,112.4) ;
%Straight Lines [id:da35133076722161927] 
\draw    (382.85,112.4) -- (405.77,112.4) ;
%Shape: Ellipse [id:dp4803646439620526] 
\draw  [color={rgb, 255:red, 208; green, 2; blue, 27 }  ,draw opacity=1 ][dash pattern={on 4.5pt off 4.5pt}] (359.5,110.66) .. controls (359.5,101.46) and (375.71,94) .. (395.71,94) .. controls (415.7,94) and (431.91,101.46) .. (431.91,110.66) .. controls (431.91,119.86) and (415.7,127.32) .. (395.71,127.32) .. controls (375.71,127.32) and (359.5,119.86) .. (359.5,110.66) -- cycle ;
%Straight Lines [id:da0482059459064883] 
\draw    (349.98,122.45) -- (287.28,157.69) ;
\draw [shift={(285.54,158.67)}, rotate = 330.65999999999997] [color={rgb, 255:red, 0; green, 0; blue, 0 }  ][line width=0.75]    (10.93,-3.29) .. controls (6.95,-1.4) and (3.31,-0.3) .. (0,0) .. controls (3.31,0.3) and (6.95,1.4) .. (10.93,3.29)   ;
%Shape: Ellipse [id:dp9168051022005741] 
\draw  [color={rgb, 255:red, 74; green, 144; blue, 226 }  ,draw opacity=1 ][dash pattern={on 4.5pt off 4.5pt}] (96.52,166.44) .. controls (96.52,157.56) and (107.33,150.36) .. (120.68,150.36) .. controls (134.02,150.36) and (144.84,157.56) .. (144.84,166.44) .. controls (144.84,175.32) and (134.02,182.52) .. (120.68,182.52) .. controls (107.33,182.52) and (96.52,175.32) .. (96.52,166.44) -- cycle ;
%Straight Lines [id:da4899629760999349] 
\draw    (61.68,167.16) -- (109.78,167.16) ;
%Straight Lines [id:da237071216077003] 
\draw    (133.6,167.16) -- (181.7,167.16) ;
%Straight Lines [id:da48899105048645297] 
\draw    (206.43,167.16) -- (254.53,167.16) ;
%Straight Lines [id:da8865929269311419] 
\draw    (284.63,182.86) -- (348.82,214.87) ;
\draw [shift={(350.61,215.76)}, rotate = 206.5] [color={rgb, 255:red, 0; green, 0; blue, 0 }  ][line width=0.75]    (10.93,-3.29) .. controls (6.95,-1.4) and (3.31,-0.3) .. (0,0) .. controls (3.31,0.3) and (6.95,1.4) .. (10.93,3.29)   ;
%Shape: Ellipse [id:dp3382562630962136] 
\draw  [color={rgb, 255:red, 74; green, 144; blue, 226 }  ,draw opacity=1 ][dash pattern={on 4.5pt off 4.5pt}] (442.17,233.04) .. controls (442.17,224.16) and (452.99,216.96) .. (466.33,216.96) .. controls (479.68,216.96) and (490.49,224.16) .. (490.49,233.04) .. controls (490.49,241.92) and (479.68,249.12) .. (466.33,249.12) .. controls (452.99,249.12) and (442.17,241.92) .. (442.17,233.04) -- cycle ;
%Straight Lines [id:da5545394334034288] 
\draw    (404.33,233.76) -- (452.43,233.76) ;
%Straight Lines [id:da6191834536787058] 
\draw    (482.55,233.76) -- (530.65,233.76) ;
%Shape: Ellipse [id:dp5008132523430777] 
\draw  [color={rgb, 255:red, 74; green, 144; blue, 226 }  ,draw opacity=1 ][dash pattern={on 4.5pt off 4.5pt}] (194.72,291.92) .. controls (194.72,283.04) and (205.53,275.84) .. (218.88,275.84) .. controls (232.22,275.84) and (243.04,283.04) .. (243.04,291.92) .. controls (243.04,300.8) and (232.22,308) .. (218.88,308) .. controls (205.53,308) and (194.72,300.8) .. (194.72,291.92) -- cycle ;
%Straight Lines [id:da3096889642194671] 
\draw    (157.88,292.64) -- (205.98,292.64) ;
%Straight Lines [id:da7863143496802387] 
\draw    (358.27,245.25) -- (295.57,280.49) ;
\draw [shift={(293.83,281.47)}, rotate = 330.65999999999997] [color={rgb, 255:red, 0; green, 0; blue, 0 }  ][line width=0.75]    (10.93,-3.29) .. controls (6.95,-1.4) and (3.31,-0.3) .. (0,0) .. controls (3.31,0.3) and (6.95,1.4) .. (10.93,3.29)   ;

% Text Node
\draw (111.92,23.96) node [anchor=north west][inner sep=0.75pt]    {$A_{4}$};
% Text Node
\draw (183.84,24.92) node [anchor=north west][inner sep=0.75pt]    {$A_{6}$};
% Text Node
\draw (41.89,24.92) node [anchor=north west][inner sep=0.75pt]    {$E_{2}$};
% Text Node
\draw (260.21,26.84) node [anchor=north west][inner sep=0.75pt]    {$D_{3}$};
% Text Node
\draw (290.22,45.49) node [anchor=north west][inner sep=0.75pt]  [font=\small,rotate=-24.91] [align=left] {{\small Transition 2}};
% Text Node
\draw (448.27,100.92) node [anchor=north west][inner sep=0.75pt]    {$A_{4}$};
% Text Node
\draw (518.39,101.88) node [anchor=north west][inner sep=0.75pt]    {$A_{6}$};
% Text Node
\draw (365.65,101.88) node [anchor=north west][inner sep=0.75pt]    {$E_{1}$};
% Text Node
\draw (594.76,103.8) node [anchor=north west][inner sep=0.75pt]    {$D_{3}$};
% Text Node
\draw (405.11,100.92) node [anchor=north west][inner sep=0.75pt]    {$A_{0}$};
% Text Node
\draw (279.47,140.86) node [anchor=north west][inner sep=0.75pt]  [font=\small,rotate=-331.83] [align=left] {{\small Transition 3}};
% Text Node
\draw (110.92,156.64) node [anchor=north west][inner sep=0.75pt]    {$A_{5}$};
% Text Node
\draw (182.84,156.64) node [anchor=north west][inner sep=0.75pt]    {$A_{6}$};
% Text Node
\draw (40.89,156.64) node [anchor=north west][inner sep=0.75pt]    {$E_{1}$};
% Text Node
\draw (259.21,158.56) node [anchor=north west][inner sep=0.75pt]    {$D_{3}$};
% Text Node
\draw (296.09,169.24) node [anchor=north west][inner sep=0.75pt]  [font=\small,rotate=-24.91] [align=left] {{\small Transition 3}};
% Text Node
\draw (453.27,223.24) node [anchor=north west][inner sep=0.75pt]    {$A_{12}$};
% Text Node
\draw (531.74,223.24) node [anchor=north west][inner sep=0.75pt]    {$D_{3}$};
% Text Node
\draw (383.54,223.24) node [anchor=north west][inner sep=0.75pt]    {$E_{1}$};
% Text Node
\draw (206.76,282.12) node [anchor=north west][inner sep=0.75pt]    {$D_{16}$};
% Text Node
\draw (137.09,282.12) node [anchor=north west][inner sep=0.75pt]    {$E_{1}$};
% Text Node
\draw (288.77,262.66) node [anchor=north west][inner sep=0.75pt]  [font=\small,rotate=-331.83] [align=left] {{\small Transition 4}};

\end{tikzpicture}
    \caption{The above graph demonstrates the algorithm to deform a theory consistent with the condition \eqref{Eq:9d_condition} to a maximally enhanced theory in Table \ref{Tab:rank17}.}
    \label{algorithm}
\end{figure}

One can first use transitions 1 and 2 to convert the enhanced theories on both ends into one of $\{C_{0~\text{or}~1},E_{n\neq 2},D_n\}$. Then, one can use transition 3 to fuse all the $A$-type points of symmetry enhancement into one. If one of the endpoints is a $D$ theory, one can use transition 4 to absorb the remaining $A$-singularity into the $D$. If one of the endpoints is $C_0$, either there is nothing in the middle, or there is just an $A_0$. In the latter case, one can use transition 5 to absorb the $A_0$ fiber into $C_0$ and change it into $C_1$. At the end of this series of transitions, the condition \ref{Eq:9d_condition} is still satisfied, and none of the transitions 1-5  can be done anymore. The only configurations that have these properties are the ones listed in Tables  \ref{Tab:rank17}, \ref{Tab:rank9}, and \ref{Tab:rank1}. Therefore, any gauge group can be enhanced to a semisimple group listed in the third column of Tables  \ref{Tab:rank17}, \ref{Tab:rank9}, and \ref{Tab:rank1}. An example of this algorithm is illustrated in Figure \ref{algorithm}. 

Note that for transitions 3-5, the brane theory encodes the data of the $U(1)$ gauge coupling through the relative position of the enhanced global symmetry points on its Coulomb branch. When the distance between these points is shrunk to zero, the appropriate adjoint vector bosons become massless, and the symmetry is enhanced. In the string theory language, this corresponds to the string states connecting the D-branes becoming massless. 

The above algorithm offers an easy way to count the number of $\mathfrak{u}(1)$'s. The number of $\mathfrak{u}(1)$'s is the number of transitions 3-5 used plus the number of $\mathfrak{u}(1)$ at the end of the algorithm! This allows us to read off the full gauge group by looking at the brane's Coulomb branch. 

There is a small loophole in the above argument that we address below. In the above algorithm we assumed that we can arbitrarily move the position of the points of symmetry enhancement to perform the transitions \ref{transitions}, as long as they are consistent with equations \eqref{coupling} and \eqref{Eq:9d_condition}. However, we might not have such a control over the brane moduli space through variations of the spacetime moduli. In the following we show that variations of bulk moduli indeed allow for such arbitrary deformations of the brane moduli space. Our argument has two steps. First we show that as long as the points do not cross, we can  arbitrarily move them around subject to the equations \eqref{coupling} and \eqref{Eq:9d_condition}. Then we show that after moving two points corresponding to one of the transitions arbitrarily close to each other, they can be fused by changing bulk moduli. 

\textbf{First step: } In Subsection \ref{Sec:7d}, we will show that by varying the vev of bulk $\mathfrak{u}(1)$ scalars, any sufficiently small movement of the points of symmetry enhancement that satisfies \eqref{coupling} and \eqref{Eq:9d_condition} is possible. Now we argue that any large movement must also be possible. Suppose we have a canonically normalized bulk modulus $\phi_{9d}$ that controls the distance between two points of symmetry enhancement. We assume that increasing $\phi_{9d}$ corresponds to bringing the two points closer to each other. The unwanted scenario happens when even by taking $\phi_{9d}$ to infinity, the points do not get arbitrarily close to each other and stop at some finite distance. As we will explain next, this cannot happen.

The 9d supersymmetry fixes the moduli space of the spacetime theory to be the moduli of the $\Gamma^{17,1}$ Narain lattice, for which all infinite distance limits are decompactification limits. Fortunately, we understand the decompactification limits for the brane theory. When two points of symmetry enhancement that cannot be fused are brought closer to one another, we get a theory that does not have any 5d UV completion and decompactifies into a 6d theory. For example, this happens when we try to fuse $A_0$ point to an $E_8$ endpoint. Therefore, the decompactification limit corresponds to the situation when two points of symmetry enhancement converge but cannot be fused. Therefore, the worrisome situation mentioned before where the points stop at some finite distance from each other never happens. This completes the first step of the argument. Now we prove the second step.

\textbf{Second step (fusibility of points): } We want to show that suppose a pair of points corresponding to one of the transitions are brought sufficiently close to each other, they can be fused. To see why, note that we can always perform the transitions in the direction of splitting a point of symmetry enhancement into two. This can be done by Higgsing the spacetime gauge symmetry. Thus, at sufficiently small distances between two points of symmetry enhancement, the relation between the canonical distance of two points of symmetry enhancement on the brane moduli and the spacetime moduli is such that the points can be fused at a finite distance of the 9d moduli space. Doing it in reverse must also be possible at finite distance of bulk moduli space.

This completes the argument that closes the loophole. We showed that any movement of the points of symmetry enhancement that satisfies equations \eqref{coupling} and \eqref{Eq:9d_condition} in addition to all of the five transitions (in both directions) are always possible through variations of the spacetime moduli. In particular, this shows that all the rank 17 gauge groups are connected by the spacetime moduli space.

\subsubsection*{Excluding 9d supergravity theories with $\mathfrak{sp}(n)$ symmetry}
Here we show that theories with $\mathfrak{sp}(n\geq2)$ symmetry can be excluded using \eqref{Eq:9d_condition} and Table~\ref{Tab:4-brane_theories}.
In order to achieve $\mathfrak{sp}(n)$ symmetry, we must take the $C_n$ theory as one of the endpoints of $S^1/\mathbb{Z}_2$.
At this time, in order to satisfy \eqref{Eq:9d_condition}
\begin{align}
&-(8+n)+\frac{c_{S^1/\mathbb{Z}_2}}{c_{A_0}}\geq0,
&&\Rightarrow
&&n\leq \frac{c_{S^1/\mathbb{Z}_2}}{c_{A_0}}-8,
\end{align}
is required, where $c_{S^1/\mathbb{Z}_2}$ is the value of $c_i$ at the other endpoint.
From here, we can see that the upper bound of $n$ for $\mathfrak{sp}(n)$ symmetry is determined by the maximum value of $c_{S^1/\mathbb{Z}_2}/c_{A_0}$ that can be taken.
Table~\ref{Tab:4-brane_theories} shows that $c_{S^1/\mathbb{Z}_2}/c_{A_0}$ is maximized in the $E_0$ theory, where the value is $9$.
This means that 
\begin{align}
n\leq \frac{c_{S^1/\mathbb{Z}_2}}{c_{A_0}}-8 \leq 1.
\label{Eq:sp_bound}\end{align}
Thus, theories with $\mathfrak{sp}(1)=\mathfrak{su}(2)$ symmetry are feasible, but theories with $\mathfrak{sp}(n\geq2)$ symmetry are not.
Note that there is nothing problematic with a supersymmetric $\mathfrak{sp}(n)$ symmetry without coupling to gravity.  Indeed the $O8^+$ orientifold in string theory realizes it on a non-compact space, where 9d gravity is decoupled, except for the running of the dilaton which leads to infinitely strong coupling at finite distance. In the string construction, the point at strong coupling can only be probed by very long strings stretching out of the $O8^+$, which have a high energy. Therefore, this singularity can be interpreted as a stringy version of the Landau pole.

Also, note that supersymmetry is essential for the conclusions reached here.
It is known that in a non-supersymmetric gravity theory, one can obtain a theory with $\mathfrak{sp}(n)$ gauge symmetry in 10 dimensions \cite{Sugimoto:1999tx}, and by compactifying it to $S^1$ one can obtain the same symmetry in 9 dimensions.
 
 \subsection{8d}\label{Sec:8d}
 In this Section, we review the argument for the 8d case discussed in \cite{Hamada:2021bbz}, and extend it to derive the SLP in 8d supergravity theories.
 
The gauge instantons in 8d are 3-branes, which are described by a 4d $\mathcal{N}=2$ rank-1 theory. 
The coupling constant is represented by an elliptic curve, and the total space is an hyperk\"{a}hler geometry.
The only known compact and connected hyperk\"{a}hler manifolds with real dimension 4 are the torus $T^4$ or K3, and only the latter produces non-Abelian symmetry. A direct application of the arguments in Section \ref{Sec:9d} shows that only K3 is relevant for theories with 16 supercharges. 
In this way, the elliptic K3 geometry of the F-theory compactification \cite{Vafa:1996xn,Morrison:1996na,Morrison:1996pp} is reconstructed from the 3-brane \cite{Hamada:2021bbz}.
Moreover, by studying the structure of the Coulomb branch for gauge instantons from the bottom-up perspective \cite{Argyres:2015ffa}, one recovers the dictionary between K3 singularities and enhanced gauge symmetries.
See \cite{Cvetic:2020kuw,Montero:2020icj,Dierigl:2020lai,Cvetic:2021sjm} for swampland constraints on the global structure of the gauge group in 8d.

In Section \ref{GD}, we have provided the argument that the rank of brane theory is one.
As remarked in the footnote there, we used the extra input from the string theory constructions to exclude the possibility of the yet-to-be-discovered SCFT.  
In 8d, we can exclude such a possibility in yet another way based on the central charge\footnote{This statement applies to 9d as well, since it is related by a simple $S^1$ compactification.}.
It is known that other gauge algebras cannot be realized \cite{Garcia-Etxebarria:2017crf,Hamada:2021bbz}, so in the following, we will only consider the case of the simply-laced gauge algebra and the $\mathfrak{sp}(n)$ gauge algebra.
Assuming that the Higgs branch is given by that of the one-instanton moduli space, the central charges $a$ and $c$ of SCFT are determined \cite{Beem:2013sza,Shimizu:2017kzs}. This is because the Higgs branch corresponds to a non-Abelian instanton of nonzero size, and the low-energy dynamics is uniquely determined by supergravity and the index theorem. 
Moreover, an upper bound to the rank can be obtained by using the relationship \cite{Shapere:2008zf} between the central charge and the scaling dimension of the Coulomb branch coordinates: \cite{Hamada:2021bbz}
\begin{align}
\text{(Rank)}\leq 4(2a-c).
\label{Eq:rank_bound}\end{align}
This can be used to rule out the existence of nontrivial SCFTs with ranks greater than one that we do not yet know about.
For $\mathfrak{sp}(n)$, the central charges are those of the trivial SCFTs, and there are no nontrivial SCFTs.
For simply-laced gauge algebras, by giving the bulk scalar a vacuum expectation value, we can always break the symmetry to $SU(2)$. 
The central charges of the interacting SCFT on the corresponding small instanton 3-brane are $a=11/24$ and $c=1/2$.
By substituting these values into \eqref{Eq:rank_bound}, this leads to $\text{(Rank)}\leq 5/3$, which means that ranks higher than one are excluded.

We will now briefly comment on the reduced rank cases, which are related to frozen singularities in F-theory.
In eight dimensions, there are three theories of rank 18, 10, and 2~\cite{Montero:2020icj}. The gauge group of the theory with rank 18 is completely reproduced from the 3-brane.
A theory of smaller rank corresponds to the case with $D_{8+n}$ frozen singularities, and the theory on the 3-brane that probes this singularity is a $S^1$ compactification of the $C_n$ theory of Table~\ref{Tab:4-brane_theories}.
A theory with rank 10 corresponds to geometry with a single frozen $D_{8+n}$ singularity, while a theory with rank 2 corresponds to geometry with two.
This mapping allows us to reproduce the gauge symmetry in theories with reduced rank too, which completes the SLP in 8d supergravity theories as well.
 
 \subsection{7d}\label{Sec:7d}
 In this Section, we apply the methodology of previous Sections to 7d theories with 16 supercharges. In 7d, the brane is magnetically charged under the 3-form field in the gravity multiplet. Therefore, the brane is a 2-brane instanton. Assuming the brane is BPS, the worldvolume theory of the 2-brane is a 3d $\mathcal{N}=4$ theory, which becomes $U(1)$ in the Coulomb branch and at low energies.

Following the black hole argument and the strong version of the cobordism conjecture reviewed in Section \ref{GD}, we can respectively argue that the moduli space is compact and connected. 
Moreover, from the  $\mathcal{N}=4$ supersymmetry, we know that the moduli space is hyperk\"{a}hler \cite{Seiberg:1996nz}. To sum up, we are led to conclude that the moduli space is a four-dimensional compact, connected hyperk\"{a}hler manifold. As mentioned above, only two such examples are known: $T^4$ and K3. The case of $T^4$ has no symmetry enhancement at any point in the moduli space and corresponds to seven-dimensional theories with 32 supercharges. We will therefore focus on the remaining case of 16 supercharges described by K3.

In principle, we can use this knowledge of moduli space to constrain the landscape of gauge theories, similar to what we did in the 9d case in Section \ref{GD}. To systematically classify the possible 7d theories of maximal rank, we will need to construct all possible SCFT's that arise at singular points in the Coulomb branch, just as we did in the 9d case and was done for the 7d case in \cite{Fraiman:2021soq}. Unlike in the 9d case in Section \ref{GD}, we do not have a systematic classification of 3d $\mathcal{N}=4$ SCFT's, so we cannot use it to produce a list of allowed singularities (including the frozen ones); this means that strictly speaking our results here are weaker than the 9d and 8d cases. What we will do instead is list the known 3d rank $\mathcal{N}=4$ SCFT's that arise at singular points in known string theory constructions,\footnote{The $M2$ brane probing M-theory on K3 was discussed in \cite{Seiberg:1996bs,Intriligator:1999cn}.} and use these singularities to reconstruct all 7d $\mathcal{N}=1$ maximal enhancement points. We can then reverse the field and string theory roles: Rather than using a field theory construction to construct all 7d $\mathcal{N}=1$ theories, we will employ the known existing constructions to \emph{predict} a classification of those rank-1 3d $\mathcal{N}=4$  SCFT's which have the instanton moduli space as the Higgs branch!

Before embarking on the classification, we must discuss a subtlety that is absent in the 8d and 9d cases. The local singularities that can arise in K3 are codimension 4. The geometry, in a neighborhood of the singularity, looks like $\mathbb{R}^4/\Gamma_g$, where $\Gamma_g$ is an ADE group. The four scalars in the worldvolume of the brane are the three scalars that live in the 3d $\mathcal{N}=4$ vector multiplet, and the dual photon. Since the low-energy theory is a sigma model into K3, there is a possibility of including a topological coupling in the brane worldvolume,
\begin{equation} \int C_{IJK} \epsilon^{\alpha\beta\gamma} \partial_\alpha X^I \partial_\beta X^J\partial_\gamma X^K= \int \pi^*(C),\label{topocop}\end{equation}
where $\{\alpha,\beta,\gamma\}$ are brane worldvolume indices, Latin uppercase indices correspond to the K3 tangent space, and the notation $\pi^*(C)$ just denotes the pullback, to the 2-brane worldvolume, of the K3 3-form $C$. A smooth K3 has no nontrivial 3-cycles, and so in such a case $C=0$. But in a singular K3, one can excise the local singularity, and the resulting space has a nontrivial linking 3-cycle of topology $S^3/\Gamma_g$. It follows that, if in a given quantum theory of gravity, we find a brane with $C\neq0$ around some singularity, it will not be possible to deform the K3 to be smooth: the corresponding singularity must be frozen. 

What we have just given is a Swampland derivation of the existence of frozen singularities, which are very familiar from F and M theory constructions \cite{deBoer:2001wca}. In particular, we have recovered their M-theory description as geometric singularities frozen by 3-form flux. In a sense, the brane perspective is telling us that any consistent 7d $\mathcal{N}=1$ theory can arise from K3 with frozen singularities, and so, it gets us tantalizingly close to the statement that M-theory is the unique quantum theory of gravity in seven dimensions. 

From the definition \eqref{topocop}, it is pretty clear that the 3-form $C$ is only defined modulo an integer, since the coupling remains the same upon shifting $C$ by a 3-form that integrates to 1 on the relevant 3-cycle. Furthermore, 3d $\mathcal{N}=4$ supersymmetry requires that the coupling is topological, which amounts to the statement that the 3-form $C$ is closed, $dC=0$.  This local condition must be true globally in the compact K3 (with singularities removed), so if there are $p=1,\ldots$ frozen singularities in K3, the holonomies $\int_{S^3/\Gamma_g^{(p)}} C$ around each of the frozen singularities must satisfy 
\begin{equation} \sum_p \int_{S^3/\Gamma_g^{(p)}} C\,\equiv\,0\, \text{mod}\, 1.\label{consistencycond}\end{equation}
The constraint \eqref{consistencycond} can also be recovered in known compactifications to seven dimensions: it just becomes the condition that there cannot be $G_4$ flux on M-theory on K3 \cite{deBoer:2001wca}.

Armed with the above, we can reproduce the list of maximal enhancements in known theories in seven dimensions \cite{deBoer:2001wca,Fraiman:2021soq}. The following table lists all known (possibly frozen) local singularities that can arise in K3, the global symmetry on the brane theory at that point (corresponding to the non-Abelian enhanced symmetry), the local geometry of the Coulomb branch near the singularity and taken from \cite{deBoer:2001wca,Tachikawa:2015wka}, and the corresponding discrete flux threading the singularity.

\begin{table}[!ht]
  \renewcommand{\arraystretch}{1.5}
    \addtolength{\tabcolsep}{3pt} 
\begin{center}
\small
\begin{tabular}{|c|c|c|}\hline \hline
  Unfrozen Algebra     &  Flux $\int C_3$  &  Frozen Algebra         \\\hline\hline
$\mathfrak{so}(2n+8)$   &  $1/2$   &  $\mathfrak{sp}(n)$     \\\hline
$\mathfrak{e}_6$   & $1/2$  &  $\mathfrak{su}(3)$      \\\hline
$\mathfrak{e}_6$   &  $1/3, 2/3$ &  $\varnothing$   \\\hline
$\mathfrak{e}_7$   & $1/2$  &  $\mathfrak{so}(7)$    \\\hline
$\mathfrak{e}_7$   & $1/3, 2/3$  &  $\mathfrak{su}(2)$   \\\hline
$\mathfrak{e}_7$   & $1/4, 3/4$  &  $\varnothing$   \\\hline
$\mathfrak{e}_8$   & $1/2$  &  $\mathfrak{f}_4$      \\\hline
$\mathfrak{e}_8$   & $1/3, 2/3$  &  $\mathfrak{g}_2$      \\\hline
$\mathfrak{e}_8$   & $1/4, 3/4$  &  $\mathfrak{su}(2)$      \\\hline
$\mathfrak{e}_8$   & $1/5, 2/5, 3/5, 4/5$  &  $\varnothing$      \\\hline
$\mathfrak{e}_8$   & $1/6, 5/6$  &  $\varnothing$      \\\hline
\end{tabular}
\caption{
The frozen singularities in the context of the compactification of M-theory on K3~\cite{deBoer:2001wca,Tachikawa:2015wka}.
}
\label{Tab:7d_frozen}
\end{center}
\end{table}

These have to be combined in all possible ways to form a K3 with singular fluxes. We list all the possibilities in the table below, matching the known list of 7d theories with reduced rank \cite{deBoer:2001wca,Kim:2019ths,Fraiman:2021soq}.
So assuming this table can be derived independently from the classification of 3d SCFT's with ${\cal N}=4$, this would complete the SLP program for supergravity theories in 7d as well.

\begin{table}[!ht]
  \renewcommand{\arraystretch}{1.2}
    \addtolength{\tabcolsep}{-1pt} 
\begin{center}
\small
\begin{tabular}{|c|c|c|c|}\hline \hline
  Rank    &  Flux $\int C_3$ & Freezing rule &  Dual description    \\\hline\hline
$11$   & $\frac{1}{2}+\frac{1}{2}$ &
\begin{tabular}{c}
$\mathfrak{so}(2n+8)\oplus\mathfrak{so}(2m+8)\to\mathfrak{sp}(n)\oplus\mathfrak{sp}(m)$ \\
$2\,\mathfrak{e}_6\to2\,\mathfrak{su}(3)$, \,$2\,\mathfrak{e}_7\to2\,\mathfrak{so}(7)$, \,$2\,\mathfrak{e}_8\to2\,\mathfrak{f}_4$ \\
$\mathfrak{so}(2n+8)\oplus \left(e_6, e_7, e_8\right)$ \\
$\to \mathfrak{sp}(n)\oplus\left(\mathfrak{su}(3),\mathfrak{su}(7),\mathfrak{f}_4\right)$\\
$\mathfrak{e}_6\oplus\mathfrak{e}_7\to\mathfrak{su}(3)\oplus\mathfrak{so}(7)$,\\
$\mathfrak{e}_6\oplus\mathfrak{e}_8\to\mathfrak{su}(3)\oplus\mathfrak{f}_4$,\,
$\mathfrak{e}_7\oplus\mathfrak{e}_8\to\mathfrak{so}(7)\oplus\mathfrak{f}_4$
\end{tabular}
&
\begin{tabular}{c}
    Hetero $\mathbb{Z}_2$ triple 	\\
    CHL string \\
    IIA $6O6^-+2O6^+$	\\
    no vector structure        \\
    F on K3$\times S^1/\mathbb{Z}_2$  
\end{tabular}     \\\hline
$7$   & $\frac{1}{3}+\frac{2}{3}$ &
\begin{tabular}{c}
$2\,\mathfrak{e}_6\to\varnothing$,\,
$2\,\mathfrak{e}_7\to2\,\mathfrak{su}(2)$,\,
$2\,\mathfrak{e}_8\to2\,\mathfrak{g}_2$  \\
$\mathfrak{e}_6\oplus\mathfrak{e}_7\to\mathfrak{su}(2)$,\,
$\mathfrak{e}_6\oplus\mathfrak{e}_8\to\mathfrak{g}_2$,\\
$\mathfrak{e}_7\oplus\mathfrak{e}_8\to\mathfrak{su}(2)\oplus\mathfrak{g}_2$
\end{tabular}
&  
\begin{tabular}{c}
Hetero $\mathbb{Z}_3$ triple \\
F on K3$\times S^1/\mathbb{Z}_3$   
\end{tabular}
\\\hline
$5$   & $\frac{1}{4}+\frac{3}{4}$ &
\begin{tabular}{c}
$2\,\mathfrak{e}_7\to\varnothing$ ,\,
$2\,\mathfrak{e}_8\to2\,\mathfrak{su}(2)$ \\
$\mathfrak{e}_7\oplus\mathfrak{e}_8\to \mathfrak{su}(2)$
\end{tabular}
&  
\begin{tabular}{c}
Hetero $\mathbb{Z}_4$ triple   \\
F on K3$\times S^1/\mathbb{Z}_4$   
\end{tabular}
\\\hline
$3$   & $\frac{1}{5}+\frac{4}{5}$ &  $2\,\mathfrak{e}_8\to\varnothing$ &  
\begin{tabular}{c}
Hetero $\mathbb{Z}_5$ triple  \\
F on K3$\times S^1/\mathbb{Z}_5$   
\end{tabular}
\\\hline
$3$   & $\frac{1}{6}+\frac{5}{6}$ & $2\,\mathfrak{e}_8\to\varnothing$  &  
\begin{tabular}{c}
Hetero $\mathbb{Z}_6$ triple \\ 
F on K3$\times S^1/\mathbb{Z}_6$   
\end{tabular}
\\\hline
$3$  & $\frac{1}{2}+\frac{1}{2}+\frac{1}{2}+\frac{1}{2}$ &
\begin{tabular}{c}
$\mathfrak{so}(8)\oplus\mathfrak{so}(2n+8)\oplus\mathfrak{so}(2m+8)\oplus\mathfrak{so}(2\ell+8)$ \\
$\to\mathfrak{sp}(n)\oplus\mathfrak{sp}(m)\oplus\mathfrak{sp}(\ell)$  
\end{tabular}
& IIA $4O6^-+4O6^+$    \\\hline
$3$  & $\frac{1}{2}+\frac{1}{2}+\frac{1}{2}+\frac{1}{2}$  & 
\begin{tabular}{c}
$\mathfrak{so}(8)\oplus\mathfrak{so}(2n+8)\oplus\mathfrak{so}(2m+8)\oplus\mathfrak{so}(2\ell+8)$ \\
$\to\mathfrak{sp}(n)\oplus\mathfrak{sp}(m)\oplus\mathfrak{sp}(\ell)$  
\end{tabular}
&     
\begin{tabular}{c}
IIA $4O6^-+4O6^+$ \\
F on $T^4\times S^1/\mathbb{Z}_2$
\end{tabular}
\\\hline
$1$  &$ \frac{1}{3}+\frac{1}{3}+\frac{1}{3} $& $3\,\mathfrak{e}_6\to \varnothing, \,2\,\mathfrak{e}_6\oplus \mathfrak{e}_7\to \mathfrak{su}(2)$  & F on $T^4\times S^1/\mathbb{Z}_3$    \\\hline
$1$  & $\frac{1}{2}+\frac{1}{4}+\frac{1}{4}$ & 
\begin{tabular}{c}
$\mathfrak{so}(8+2n)\oplus \mathfrak{e}_7\oplus \left(\mathfrak{e}_{7}, \mathfrak{e}_{8}\right)$ \\
$\to \mathfrak{sp}(n)\oplus\left(\varnothing,\mathfrak{su}(2)\right)$
\end{tabular}
& F on $T^4\times S^1/\mathbb{Z}_4$
\\\hline
$1$  & $\frac{1}{2}+\frac{1}{3}+\frac{1}{6}$ & 
$\mathfrak{so}(2n+8)\oplus \left(\mathfrak{e}_6,\mathfrak{e}_7\right) \oplus \mathfrak{e}_8\to \mathfrak{sp}(n)\oplus\left(\varnothing,\mathfrak{su}(2)\right)$  
& F on $T^4\times S^1/\mathbb{Z}_6$    \\\hline
\end{tabular}
\caption{
List of 7d theories with reduced rank~\cite{deBoer:2001wca,Kim:2019ths,Fraiman:2021soq}.
The reduced rank theories are obtained by putting the flux in Table~\ref{Tab:7d_frozen}, where the total flux must vanish mod $1$ in the compact manifold. The gauge algebra is obtained by the replacement of the maximal rank theories (the list of maximally enhanced gauge algebra is given in \cite{Fraiman:2021soq}).
The number of inequivalent rank-3 theories is not entirely certain. 
It may be four rather than three (see footnote 22 in \cite{deBoer:2001wca}).
Applying the freezing rule to the list provided in \cite{Fraiman:2021soq}, we see that the gauge algebra of all rank-1 theories can be enhanced to $\mathfrak{su}(2)$. The only maximal enhanced gauge algebra of rank-3 theories corresponding to $4O6^-+4O6^+$ is $3\,\mathfrak{su}(2)$.
The maximal gauge algebra of the other cases is given in \cite{Fraiman:2021soq}.
}
\label{Tab:7d_theories}
\end{center}
\end{table}

As we have seen so far, much information about the spacetime theory is encoded in the small instanton moduli space. For example, the gauge group of the spacetime theory is related to the singularities of the brane moduli space. An ambitious improvement of this relationship would be to understand how the Coulomb branch of the brane moduli deforms by changing the spacetime moduli. In fact, in 7d, we can make an elegant connection between the geometry of the brane moduli space and the spacetime moduli. 

Take a 2-cycle in the small instanton moduli space and consider a skyrmion where the brane moduli wrap around the 2-cycle on the spatial slices of the brane. Suppose we can localize the 2+1 dimensional skyrmion so that it pinches off from the brane worldvolume. The pinched-off skyrmion is a 0+1 spacetime particle. Note that the mass of the scalar field corresponding to this particle controls the size of the 2-cycle. Therefore, we can locally control the complex structure of the K3 moduli space subject to the frozen singularities by changing the periods of K3 through varying the spacetime moduli. This  establishes a direct relationship between the brane moduli space and spacetime moduli using only field theory. 

The above result holds for higher dimensions as well. For example, take the 9d theory and compactify it on a $T^2$ down to 7d. The points of symmetry enhancements on the brane Coulomb branch map to singularities of K3. Therefore, we can locally move the location of the singularities subject to the K3 geometry by changing the spacetime moduli. If we decompactify one $S^1$, the path in the 7d moduli space lifts to a path in the 8d moduli space, which moves the location of the singular fibers on the brane moduli space subject to the geometry of the elliptic K3. Suppose we further decompactify the extra $S^1$. In that case, the path lifts to a local movement of the points of symmetry enhancement on the brane Coulomb branch subject to the single valuedness of gauge coupling $g$ in \eqref{coupling} and $\sum_i c_i=0$ from \eqref{Eq:9d_condition}. 

If the spacetime gauge group of the 9d theory differs from the ones listed in Tables \ref{Tab:rank17}, \ref{Tab:rank9}, and \ref{Tab:rank1}, two things happen simoulaniously:

\begin{enumerate}
\item Given that the entries of the tables have maximal semisimple algebras, the gauge algebra must be a subalgebra of one of the entries with an additional $\mathfrak{u}(1)$ factor.  

\item Since the entries of the tables are the only configurations where the relative positions of the points of symmetry enhancement are completely frozen, the relative position between at least two of the points of symmetry enhancement must be tunable.
\end{enumerate}

Therefore, we conclude the relative position between two points of symmetry enhancements is tunable if and only if the gauge algebra has an additional $\mathfrak{u}(1)$ factor. In other words, the Coulomb branch of the brane moduli space is sensitive to the scalars of the vector multiplets corresponding to the Abelian $\mathfrak{u}(1)$'s.

 \subsection{6d} 
In 6d, there are two types of theories with 16 supersymmetries: chiral ($\mathcal{N}=(2,0)$) and non-chiral ($\mathcal{N}=(1,1)$). 
Of these, the chiral $\mathcal{N}=(2,0)$ theory is known to have such a strong restriction that the massless spectrum is determined by symmetry alone \cite{Townsend:1983xt}, so we will consider the non-chiral $\mathcal{N}=(1,1)$ theory.
 
The initial analysis of the 6d theories parallels that of the 7d. The brane is $(1+1)$ dimensional and has $\mathcal{N}=(4,4)$ supersymmetry by studying the gauge instanton solution in the bulk 6d theory. 
 From a combination of the strong cobordism conjecture and the ADHM construction, we find that the moduli space of the 1-brane is a connected two-dimensional complex manifold. 
Moreover, from the black hole argument reviewed in \ref{GD}, we know that the moduli space must be compact. 
Contrary to the 7d case, the $\mathcal{N}=(4,4)$ supersymmetry does not lead to an hyperk\"{a}hler manifold as the target space of the sigma model in general \cite{Gates:1984nk,Gibbons:1997iy,Diaconescu:1997gu}.
Only when there is an extra $U(1)$ isometry does the target space becomes hyperk\"{a}hler.

If we assume an additional $U(1)$ isometry, these facts collectively narrow down the possibilities to either K3 or $T^4$. Like the 7d case, $T^4$ corresponds to theories with $32$ supercharges. Thus, we conclude that the moduli space of the 1-brane is K3.
As in the cases of the other dimensions, by classifying 2d $\mathcal{N}=(4,4)$ theories, the possible gauge algebras of 6d theories are obtained.
It would be interesting to complete the classification.

In fact, without assuming $U(1)$ isometry, we can see the appearance of K3 geometry from another argument. 
Since the 1-brane we are considering has rank-1 and $\mathcal{N}=(4,4)$, the worldsheet theory has four scalars and four fermions, and the total central charge is $c=6$.
In \cite{Eguchi:1988vra}, by calculating the elliptic genus of the $\mathcal{N}=(4,4)$ theory with $c=6$, it is shown that this theory is interpreted as a string propagating on K3\footnote{As a technical assumption, it is required that the massless representations with isospin $l=0$ and $l=1/2$ do not mix \cite{Eguchi:1988vra}. This rules out torus compactifications.}.
In this sense, we can reconstruct the K3 geometry as the target space of the sigma model.
Note that this argument can only be applied to smooth K3 and not to singular K3 with frozen singularities.

The arguments above indicate that the geometry is morally K3, but we can not rule out the possibility that a there is a $\mathcal{N}=(4,4)$ SCFT which is different from the SCFT we get from K3 (despite that fact that at least in the smooth case it must have the same elliptic genus as K3)\footnote{It would be interesting to study the patterns of frozen singularities which will appear in \cite{Fraiman-Parra-to-appear} from this point of view.}.
Modulo the assumption that all the $\mathcal{N}=(4,4)$ SCFT's with $c=6$ and compact target spaces are somehow equivalent (e.g., T-duality) to the SCFT with K3 target space (possibly with singularities), the SLP is valid.

\section{Conclusions} \label{Sec:conclusion}
In this paper, we have focused on theories with 16 supercharges and have used the existence of universal brane, which also corresponds to the small instanton limit for non-Abelian gauge fields, to complete the SLP for supersymmetric theories in dimensions above 7.  In the 7-dimensional case, the SLP is more or less complete for supergravity theories, modulo a conjecture we have made about the structure of certain 3d SCFT's with ${\cal N}=4$ supersymmetry.   We have also made progress in establishing the SLP for theories with 16 supercharges in 6d.
The main tool has been the combination of local knowledge about the structure of branes associated with small gauge instantons with the global knowledge of compactness and connectedness of the Coulomb branch of the brane theory.  In this way, we have been able to reconstruct the internal geometry of string compactifications directly from the EFT coupled to supergravity.

One aspect of the present work which is left for future work is to bring to life the meaning of the coincidence of the brane moduli with the internal stringy geometry.  In particular, we need to show a relation between the KK towers and the eigenspectrum of Laplacian on the brane moduli.  Progress in this direction will be reported elsewhere \cite{Upcoming}.

\section*{Acknowledgments}
We thank Mariana Gra\~{n}a, H\'{e}ctor Parra de Freitas, Bernardo Fraiman, Sergio Cecotti, and Houri-Christina Tarazi for useful discussions and comments. We thank the Summer Program of the Simons Center for Geometry and Physics for kind hospitality. The work of AB, MM, and CV is supported by a grant from the Simons Foundation (602883, CV) and by the NSF grant PHY-2013858. The work of YH is supported by JSPS Overseas Research Fellowships.

\bibliographystyle{JHEP}
\bibliography{Bibliography}

\end{document}